\documentclass[9pt,article,twoside]{rmaa-rho-class/rmaa-rho}
\RMxAAtemplatetype{\RMxAA} % Select your article type
\usepackage{natbib}
\usepackage{hyperref}
\usepackage[justification=justified,
   format=plain]{caption}

\newcommand{\nii}{[\ION{N}{ii}]}

\newcommand{\oiii}{[\ION{O}{iii}]}

\newcommand{\siii}{[\ION{S}{iii}]}

\DeclareRobustCommand{\ION}[2]{%
\relax\ifmmode
\ifx\testbx\f@series
{\mathbf{#1\,\mathsc{#2}}}\else
{\mathrm{#1\,\mathsc{#2}}}\fi
\else\textup{#1\,{\mdseries\textsc{#2}}}%
\fi}

\newcommand{\ha}{$\rm{H}\alpha$}
\newcommand{\hb}{$\rm{H}\beta$}
\newcommand{\hd}{$\rm{H}\delta$}
\newcommand{\hg}{$\rm{H}\gamma$}
\newcommand{\he}{$\rm{H}\epsilon$}

\newcommand{\Ha}{$\rm{H}\alpha$}
\newcommand{\Hb}{$\rm{H}\beta$}
\newcommand{\Hg}{$\rm{H}\gamma$}
\newcommand{\He}{$\rm{H}\epsilon$}

\newcommand{\pyf}{\texttt{pyFIT3D}}

\vol{61}
\pages{30-36}
\thisyear{2025}
\doi{\href{https://www.astroscu.unam.mx/rmaa/RMxAA..XX-X}{https://www.astroscu.unam.mx/rmaa/RMxAA..XX-X}}

\title{SDSS-V Local Volume Mapper (LVM): Helix Nebula public data, Data Analysis Pipeline data products}

\newcommand{\Carnegie}{Observatories of the Carnegie Institution for Science, 813 Santa Barbara Street, Pasadena, CA 91101, USA}

\newcommand{\HD}{Astronomisches Rechen-Institut, Zentrum f\"{u}r Astronomie der Universit\"{a}t Heidelberg, M\"{o}nchhofstra\ss e 12-14, D-69120 Heidelberg, Germany}

\newcommand{\MPIA}{Max-Planck-Institut f\"{u}r Astronomie, K\"{o}nigstuhl 17, D-69117, Heidelberg, Germany}

\newcommand{\UChile}{Departamento de Astronom\'{i}a, Universidad de Chile, Camino del Observatorio 1515, Las Condes, Santiago, Chile}

\newcommand{\STScI}{Space Telescope Science Institute, 3700 San Martin Drive, Baltimore, MD 21218, USA}

\newcommand{\UCatolica}{Instituto de Astronom\'ia, Universidad Cat\'olica del Norte, Av. Angamos 0610, Antofagasta, Chile}

\newcommand{\UT}{McDonald Observatory, The University of Texas at Austin, 1 University Station, Austin, TX 78712-0259, USA}

\newcommand{\NAOC}{Chinese Academy of Sciences South America Center for Astronomy, National Astronomical Observatories, CAS, Beijing 100101, China}

\newcommand{\CASA}{Center for Astrophysics and Space Astronomy, University of Colorado, 389 UCB, Boulder, CO 80309-0389, USA}

\newcommand{\UDP}{Instituto de Estudios Astrof\'isicos, Facultad de Ingenier\'ia y Ciencias, Universidad Diego Portales, Av. Ej\'ercito Libertador 441, Santiago, Chile}

\newcommand{\UNAMCU}{Universidad Nacional Aut\'onoma de M\'exico, Instituto de Astronom\'ia, AP 70-264, CDMX 04510, M\'exico}

\newcommand{\IRyA}{Instituto de Radioastronomía y Astrofísica, Universidad Nacional Autónoma de México, Morelia 58089, Michoacán, Mexico; Facultad de Ciencias de la Tierra y el Espacio, Universidad Autónoma de Sinaloa, Josefa Ortiz de Domínguez S/N, Culiacán 80040, Sin., Mexico}

\newcommand{\UNAMe}{Universidad Nacional Aut\'onoma de M\'exico, Instituto de Astronom\'ia, AP 106, Ensenada 22800, BC, M\'exico}

\newcommand{\UNAMICF}{Universidad Nacional Autónoma de México, Inst. de Ciencias F\'isicas, Av. Universidad S/N, Chamilpa, 62200 Cuernavaca, Morelos, México}

\newcommand{\IAC}{Instituto de Astrof\'\i sica de Canarias, La Laguna, Tenerife, E-38200, Spain}

\newcommand{\UNSER}{Department of Astronomy, Universidad de La Serena, Av. Raul Bitran 1302, La Serena, Chile}

\newcommand{\UCatA}{Universidad Cat\'olica del Norte, N\'ucleo UCN en Arqueolog\'ia Gal\'actica - Inst. de Astronom\'ia, Av. Angamos 0610, Antofagasta, Chile}

\newcommand{\UCatB}{Universidad Cat\'olica del Norte, Departamento de Ingenier\'ia de Sistemas y Computaci\'on, Av. Angamos 0610, Antofagasta, Chile}

\newcommand{\NYUAbu}{New York University Abu Dhabi, PO Box 129188, Abu Dhabi, UAE}
\newcommand{\CASS}{Center for Astrophysics and Space Science (CASS), New York University Abu Dhabi, PO Box 129188, Abu Dhabi, UAE}
\newcommand{\UMoscow}{Sternberg Astronomical Institute, Lomonosov Moscow State University, Universitetskij pr., 13,  Moscow, 119234, Russia}

\author[1,2,$\dagger$]{S.~F.~S\'anchez \orcidlink{0000-0001-6444-9307}}
\author[3]{J.E. M\'endez-Delgado\orcidlink{0000-0002-6972-6411}}
\author[4]{A. Mej\'\i a-Narv\'aez}
\author[1]{C. Rom\'an-Zu\~niga \orcidlink{0000-0001-8600-4798}}
\author[5]{O.~V.~Egorov}
\author[6]{C. Morisset \orcidlink{0000-0001-5801-6724}}
\author[7]{N. Drory}
\author[8,4]{G. A. Blanc\orcidlink{0000-0003-4218-3944}}
\author[5]{K. Kreckel}
\author[9]{E. J. Johnston}
\author[10,11,12]{Ivan Yu. Katkov\orcidlink{0000-0002-6425-6879}}
\author[13]{A. Roman Lopes\orcidlink{0000-0002-1379-4204}}
\author[1]{M. A. Villa-Durango\orcidlink{0009-0008-1605-4771}}
\author[3]{H. Ibarra-Medel}
\author[14]{H.-W. Rix}
\author[1]{R. de J. Zermeño}
\author[15,16]{J. G. Fern\'andez Trincado}
\author[4]{A. Singh}
\author[17,18]{P. García}
\author[19]{G. S. Stringfellow}
\author[1]{L. Sabin}
\author[20]{J. Toal\'a}
\author[20]{R. Orozco Duarte\orcidlink{0000-0003-2193-3005}}
\author[21]{A. M. Jones\orcidlink{0000-0002-2262-8240}}

\affil[1]{\UNAMe}
\affil[2]{\IAC}
\affil[3]{\UNAMCU}
\affil[4]{\UChile}
\affil[5]{\HD}
\affil[6]{\UNAMICF}
\affil[7]{\UT}
\affil[8]{\Carnegie}
\affil[9]{\UDP}
\affil[10]{\NYUAbu}
\affil[11]{\CASS}
\affil[12]{\UMoscow}
\affil[13]{\UNSER}
\affil[14]{\MPIA}
\affil[15]{\UCatA}
\affil[16]{\UCatB}
\affil[17]{\NAOC}
\affil[18]{\UCatolica}
\affil[19]{\CASA}
\affil[20]{\IRyA}
\affil[21]{\STScI}

\affil[$\dagger$]{This project is part of the SDSS collaboration}

\leadauthor{Sánchez S.F. et al.}
\smalltitle{Helix Nebula DAP dataproducts}

\corres{Sebastián F. Sánchez}
\email{sfsanchez@astro.unam.mx}

\received{xxxxx,  2025}
\accepted{xxxx, 2026}
\license{Texto de la licencia aquí}

\setbool{rho-abstract}{true} % Set false to hide the abstract
\setbool{rho-resumen}{true} % Set false to hide the abstract

\begin{abstract}
We present a spatially resolved spectroscopic analysis of the Helix Nebula (NGC\,7293) using data from the SDSS-V Local Volume Mapper (LVM), by applying the recently developed LVM Data Analysis Pipeline (LVM-DAP). Covering the full optical range (3600-9800\AA) over a contiguous $\sim0.5^\circ$ field, the LVM data provide the first hexagonally sampled, wide-field emission-line maps of all major ionic species in this archetypal planetary nebula. The resulting flux, kinematic, and line-ratio maps reveal the well-known ionization stratification of the nebula, from the compact He$^{++}$ core to the bright [\ION{O}{iii}] ring and the extended low-ionization envelope, enabling a detailed comparison with classical aperture spectroscopy. Owing to the sensitivity and uniform spatial sampling of the LVM, numerous faint auroral and diagnostic lines are detected across the nebula, including [\ION{O}{iii}]\,4363, [\ION{N}{ii}]\,5755, and He\,\textsc{i} lines, allowing precise measurements of weak-line morphology. The derived radial trends confirm the remarkably low dust content and the overall homogeneity of electron temperature and density across the main ring. Ionized-gas kinematics traced by \ha\ further support the scenario of a slowly expanding, limb-brightened shell consistent with previous studies. This work demonstrates the diagnostic power of LVM spectroscopy for extended nebulae and highlights its capability to recover both global and spatially resolved physical conditions across complex ionized structures.
\
\end{abstract}


\begin{resumen}
Presentamos un análisis espectroscópico espacialmente resuelto de la Nebulosa de la Hélice (NGC,7293) con datos del SDSS-V Local Volume Mapper (LVM), procesados con el LVM Data Analysis Pipeline (LVM-DAP). Cubriendo 3600--9800 \AA\ sobre un campo contiguo de $\sim0.5^\circ$, el LVM provee de los primeros mapas de líneas de emisión de campo amplio, muestreados hexagonalmente, de las principales especies iónicas en esta nebulosa planetaria arquetípica. Los mapas de flujo, cinemática y cocientes de líneas revelan la estratificación de ionización desde el núcleo compacto de He$^{++}$ hasta el anillo brillante de [\ION{O}{iii}] y la envoltura extendida de baja ionización, permitiendo una comparación detallada con datos de espectroscopía clásica de apertura. Gracias a la sensibilidad y al muestreo espacial uniforme del LVM, se detectan en toda la nebulosa numerosas líneas  débiles, incluyendo [\ION{O}{iii}],4363, [\ION{N}{ii}],5755 y He,\textsc{i}, lo que permite medir con precisión la morfología de esta nebulosa. Las tendencias radiales confirman el contenido de polvo notablemente bajo y la homogeneidad global de la temperatura y densidad electrónicas en el anillo principal. La cinemática del gas ionizado trazada por \ha\ apoya el escenario de una envoltura en lenta expansión realzada en los bordes, en acuerdo con estudios previos. Este trabajo demuestra el poder diagnóstico de la espectroscopía obtenida por el LVM en nebulosas extendidas y su capacidad para recuperar condiciones físicas globales y espacialmente resueltas en estructuras ionizadas complejas.
  \end{resumen}

\begin{document}

\maketitle
\pagestyle{fancy}\thispagestyle{firststyle}

%----------------------------------------------------------

\section{INTRODUCTION}
\label{sec:intro}

%%%%%%%%%%%%%%%%%%%%%%%%%%%%%%%%%%%%%%%%%%%%%%%%%%%%%%%%%%%%%%%%%%%%%%%5
\begin{figure}[t]
        \centering                
        \includegraphics[width=0.47\textwidth,clip,trim=35 25 5 5]{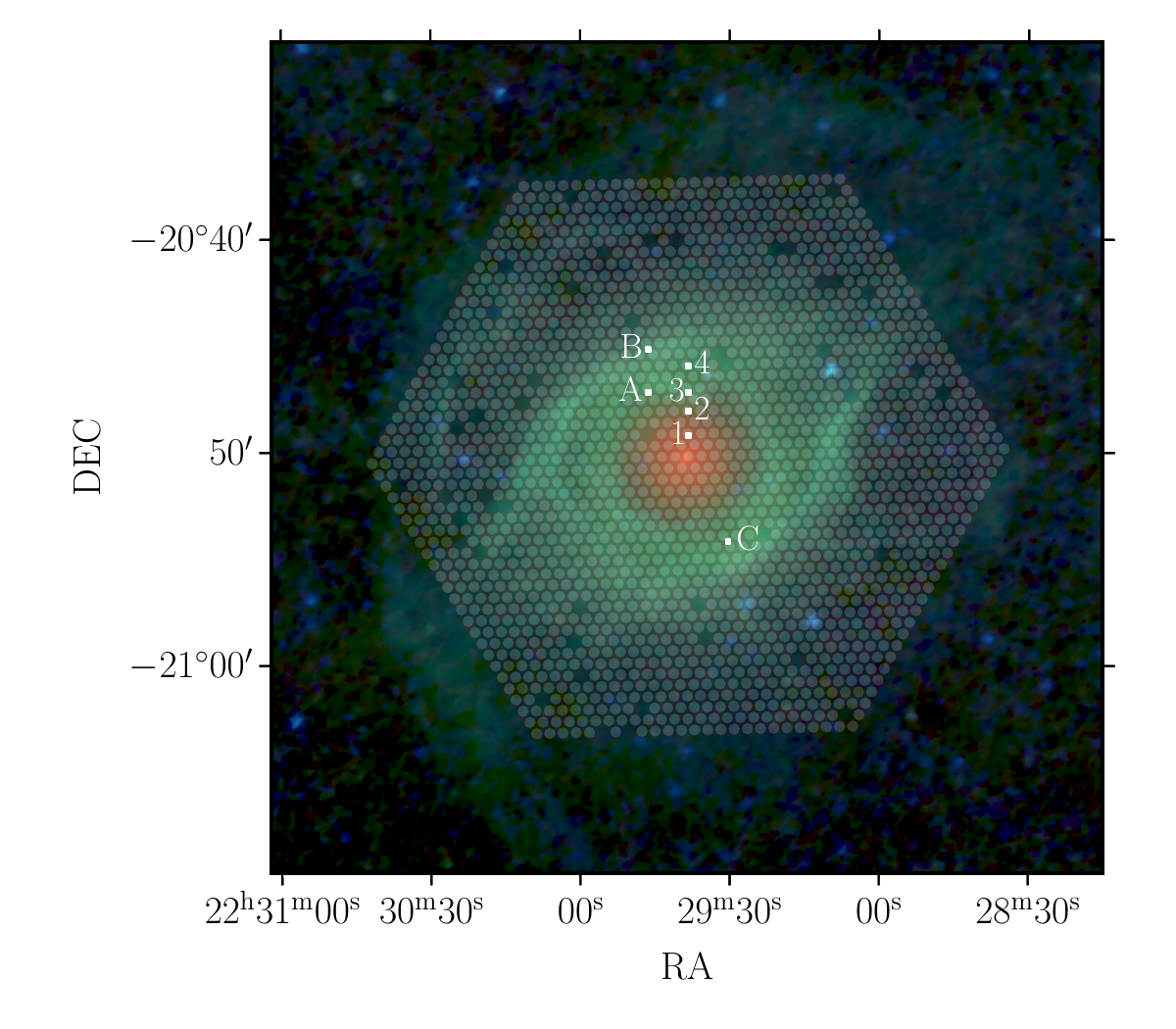}
        \captionof{figure}[]{Color image created combining the WISE W4 (22$\mu$m, red), W3 (12$\mu$, green) and W2 (4.6$\mu$m, blue) band images covering $\sim$0.65\textdegree\ size centered in { Helix Nebula} \citep[i.e., a similar FoV of the images explored by][]{odell98}. The foot-print of the LVM science IFU fibers included in the exposure delivered in the SDSS DR19 is represented by a set of semi-transparent white circles. The white points indicate the location of previous regions explored in the literature. The regions discussed by \citet{odell98} are represented by numbers: (1) 'middle', (2) 'transition', (3) 'ring' and (4) 'arc'. Finally, the regions discussed by \citet{henry99} are labeled with the same letters adopted in that article.}
        \label{fig:wise}
    \end{figure}

%\RMxAAstart{T}BW

%%%%%%%%%%%%%%%%%%%%%%%%%%%%%%%%%%%%%%%%%%%%%%%%%%%%%%%%%%%%%%%%%%%%%%%5
\begin{figure*}
        \centering                
        \includegraphics[width=0.89\textwidth,clip,trim=0 35 5 5]{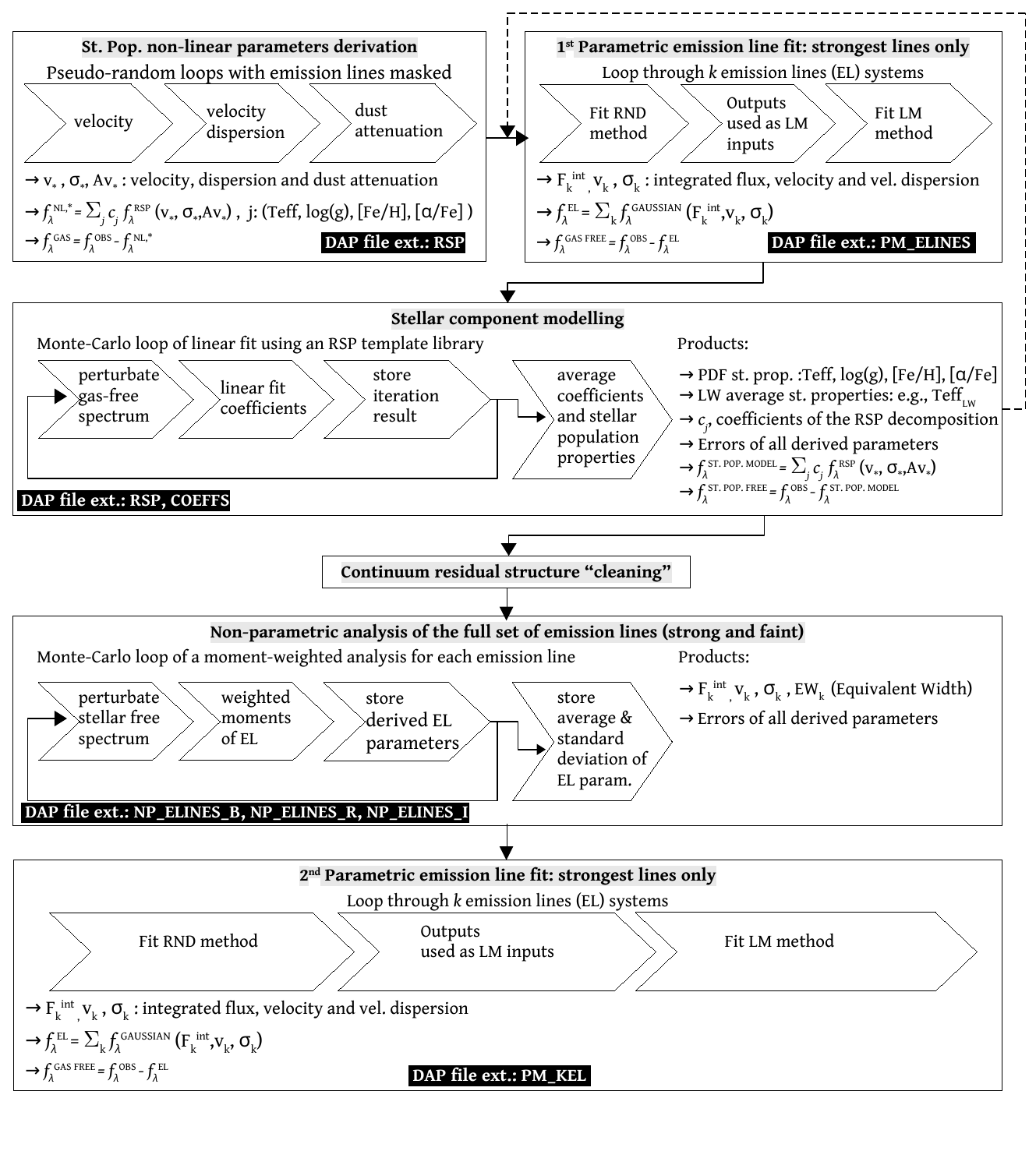}
        \captionof{figure}[]{Updated scheme of the LVM-DAP analysis flow for a single fiber spectrum, including the main procedures: (i) derivation of the non-linear parameters of the stellar spectrum (v$_\star$, $\sigma_\star$ and A$_{\rm V,\star}$), (ii) parametric derivation of the properties of the ionized gas emission lines { (EL), including the flux intensity (f$_{\rm EL}$), velocity (v$_{\rm EL}$) and velocity dispersion ($\sigma_{\rm EL}$)}, (iii) stellar component synthesis, i.e., decomposition into a set of RSP templates, and generating a model of the stellar spectrum, (iv) non-parametric derivation of the properties of the emission lines, including the equivalent width for each emission line (EW$_{\rm EL}$), and (v) a re-evaluation of the parametric derivation of the emission lines. The last two analysis were performed over a so-called gas-pure spectrum, i.e., the residual of subtracting the stellar component model from the original spectrum. RND and LM stands for the two methods included in \pyf\ to fit the parametric models to the EL, as explained in the text. Black boxes indicate in which extension of the DAP data products files the analysis of each module is stored.}
        \label{fig:scheme}
      \end{figure*}

Planetary nebulae (PNe) represent the late evolutionary stages of low-
and intermediate-mass stars (1--8~M$_\odot$), tracing the transition
from the asymptotic giant branch (AGB) to the white dwarf
phase. During this phase, material ejected from the stellar envelope
becomes ionized by the increasingly hot central star, giving rise to
rich emission-line spectra that provide key diagnostics of physical
conditions, ionization structure, and chemical enrichment. Among the
nearest and best-studied examples is the Helix Nebula
  (NGC~7293), whose proximity ($\sim$215~pc; \citealt{harris07}) and
large angular size ($\sim$16$'$~$\times$~12$'$) make it an ideal
benchmark for testing spatially resolved photoionization models and plasma
diagnostics at high sensitivity.

The Helix has been the subject of extensive spectroscopic
investigations over the past five decades, covering the optical,
ultraviolet, infrared, and radio regimes. Early studies by
\citet{warn75} and \citet{hawl98} established its basic ionization
stratification and provided the first integrated measurements of key
diagnostic lines such as [\ION{O}{iii}]5007, [\ION{N}{ii}]6583, and
[\ION{S}{ii}]6716,6731, relative to
\hb . These efforts revealed a relatively low-ionization nebula
with strong [\ION{N}{ii}] and [\ION{O}{i}] emission in the outer
zones and a prominent \ION{O}{iii} emission in the central regions, reflecting a significant contribution from partially ionized
gas.

Subsequently, \citet{leene87} conducted long-slit optical spectroscopy across
several positions, { demonstrating the strong} radial variation of
excitation conditions and elemental abundances. Their results
confirmed that helium and oxygen abundances are relatively uniform,
while nitrogen and sulfur exhibit gradients consistent with ionization
stratification rather than true chemical inhomogeneity\footnote{It is worth noticing that this may be
due to the use of ionization correction factors to account for the unobserved ions, which may
depend on the ionization degree.}. The improved
spatial sampling of \citet{odell98}, using narrow-band imaging and
region-integrated spectra, revealed substantial structural complexity
and confirmed the coexistence of multiple ionization regimes within
the nebula, associated with the bright inner ring and fainter outer
arcs.

A more quantitative abundance analysis was later provided by
\citet{henry99}, who presented high-quality optical line ratios from
three distinct slit positions. Their work remains the reference
dataset for chemical abundances in NGC~7293, establishing
characteristic values of 12~+~log(O/H)~$\approx$~8.7 and
He/H~$\approx$~0.11, comparable to solar abundances but with mild
nitrogen enhancement typical of Type~II PNe. These studies
collectively defined the observational framework against which new,
spatially resolved surveys may be compared.

Beyond optical spectroscopy, a wide variety of multi-wavelength
observations have further characterized the Helix
Nebula. \citet{bublitz22} analyzed the molecular and CO chemistry in
the outer envelope, tracing the survival of dense knots and their
connection to photodissociation regions (PDRs). \citet{etxaluze14}
presented \textit{Herschel} submillimeter and far-infrared spectra,
detecting lines such as OH$^+$, CO, and [N\,\textsc{ii}]~205~$\mu$m,
which probe the transition from ionized to molecular gas. The series
of studies by \citet{meaburn92, meaburn05, meaburn08} and
\citet{walshmeaburn87} focused on kinematics and line profiles
(\ha, [\ION{N}{ii}]), providing detailed insight into the
three-dimensional morphology and velocity field of the nebula, { albeit}
without comprehensive line-intensity catalogs. More recent
works, such as \citet{iskandarli24} and
\citet{estradadorado25}, have revisited the properties of the hot
central white dwarf and its UV spectrum, while
\citet{andriantsaralaza20} explored related modeling approaches to
planetary nebula evolution. Together, these studies emphasize that the
Helix remains one of the most comprehensively observed planetary
nebulae across the electromagnetic spectrum, yet systematic,
wide-field optical mapping of its ionization structure is still
limited.

Recent wide-field spectroscopic facilities now enable complete,
contiguous mapping of extended nebulae with high spatial and spectral
fidelity. The Local Volume Mapper (LVM), one of the SDSS-V surveys
\citep{juna25, drory25}, provides an unprecedented opportunity to
revisit nearby planetary nebulae such as the Helix with integral field
spectroscopy over degree-scale fields. In this work, we present an
analysis of the Local Volume Mapper of the Helix Nebula
included in the recent SDSS-V 19th Data Release
\citep[DR19][]{DR19}, emphasizing the spatially integrated
emission-line ratios and their comparison with canonical optical
datasets. By benchmarking the LVM measurements against legacy studies from \citet{warn75} to \citet{odell98} and \citet{henry99} we evaluate
both the consistency of derived physical conditions and the extent to
which large-scale integral field observations can capture the complex
ionization structure of nearby planetary nebulae.

We should highlight the uniqueness of the LVM IFU, that with an angular size of $\sim$0.5$\arcmin$ covers almost the { Helix Nebula} in a single pointing sampling it with $\sim$1750 fibers of 35.5$\arcsec$/diameter ($\sim$0.037 pc at the distance of the object). Furthermore, we should note that the analyzed frame is just one of the several exposures that the LVM is taking on this target, which, when combined, will provide much deeper (higher-S/N) data that will be delivered to the public in future data releases. 

The structure of this article is as follows: in Section \ref{sec:data} we describe the LVM observations and data-reduction
procedures. Section \ref{sec:ana} describes the analysis performed, including a brief description of the LVM data analysis pipeline (LVM-DAP), placing an emphasis on the modifications adopted since the previous version \citep{lvmdap}. The implemented 2nd order sky correction and the selection of a final sample of high-quality emission lines (golden sample) are described in Sections \ref{sec:sky} and \ref{sec:good}. The main results of this analysis is presented in Section \ref{sec:res}, including a description of (i) the distributed dataproducts (Section  \ref{sec:res_dp}), (ii) the integrated and spatial resolved flux intensities of golden sample emission lines (Sections \ref{sec:res_int} and \ref{sec:res_map}, respectively), (iii) the radial distribution of the absolute flux intensities  (Section \ref{sec:res_rad}) and their relative values with respect to \hb (Section \ref{sec:norm}), (iv) the ionized gas kinematics (Section \ref{sec:res_kin}), and (v) the average properties of the stellar population captured within the field-of-view (Section \ref{sec:st}). The results are discussed in Section \ref{sec:dis} and the main conclusions of this study are presented in Section \ref{sec:con}.

\section{DATA}
\label{sec:data}

The dataset analyzed in this work corresponds to a single pointing of
the Helix Nebula (NGC~7293) obtained by the LVM during its
commissioning phase at Las Campanas Observatory, Chile, on the night of September 4, 2023. This pointing, identified by the codes
MJD = 60191 / Exposure= 4297, was part of the 19th public Data Release
\citep[DR19][]{DR19} of the SDSS-V project \citep[][]{juna25}
\footnote{\url{https://www.sdss.org/dr19/lvm/}}.

The instrumental configuration of the LVM comprises an IFU array fed by four Alt-Az telescopes, one dedicated to science data acquisition (T1, feeds the ultra-wide IFU, 1801 fibers), two monitoring the sky towards East and West, (T2 and T3, 59 and 60 fibers, respectively) and one more for observing spectrophotometric calibration stars (T4, 24 fibers). The IFU array consists of DESI-110 like spectrographs covering the 3600 to 9800~{ \AA\ wavelength} range with a spectral resolution of $R \sim 4000$ at \ha. Each spectrograph employs a dichroic system to split each fiber beam into three wavelength channels: (b) blue, from 3600 to 5800 \AA; (r) red, from 5750 to 7570 \AA; and (z) infrared, from 7520 to 9800 \AA. The ultra-wide IFU provides detailed spectral and spatial coverage across the survey targets.

%This system
%feeds three DESI-like spectrographs that cover the wavelength range
%from 3600 to 9800~\AA\ with a spectral resolution of $R \sim 4000$ at
%H$\alpha$. Each spectrograph employs a dichroic system that divides
%the light from each fiber into three wavelength channels: blue (b),
%from 3600 to 5800~\AA; red (r) , from 5750 to 7570~\AA; and infrared (z), from 7520 to %9800~\AA. The extensive field of view of the
%LVM ultra-wide IFU provides detailed spectral and spatial coverage
%across the survey targets, including the Milky Way, the Magellanic
%Clouds, and a representative sample of local-volume galaxies.

The LVM survey Data Reduction Pipeline (DRP; Mejía-Narváez et al.,
in prep.) follows the procedures described by \citet{sanchez06a}. The
pipeline is based on the {\sc py3D} reduction package, originally
developed for the CALIFA survey \citep[][]{dr1}. Initially implemented
in Python~2.7, the code has been upgraded to Python~3.11 and
supplemented with dedicated routines specifically designed to address
the unique characteristics and requirements of the LVM dataset.
The current dataset was reduced using version 1.1.1 of the DRP.

The primary processing by the DRP consists of 9 stages: (i) Preprocessing of raw data frames; (ii) identification and tracing of fiber spectra on the CCDs; (iii) extraction of spectra, (iv) wavelength calibration and linear resampling of spectra; (v) differential correction for fiber-to-fiber transmission variations; (vi) flux calibration using data from T4; (vii) combination of the three spectral channels into single spectra; (viii) night sky background subtraction using data from T2 and T3, and (ix) computation of astrometric solution using data from guiding cameras. Throughout the entire reduction process, uncertainties are consistently propagated at every
stage.

%The primary processing steps executed by the LVM-DRP comprise: (i)
%preprocessing of the raw frames, in which the readouts from multiple
%amplifiers are merged into a single image for each spectrograph
%channel (b, r, and z), resulting in nine individual frames, followed
%by bias subtraction, gain correction, and the detection and masking of
%cosmic rays; (ii) identification and tracing of the fiber spectra on
%each CCD for all spectrographs, including the measurement of the FWHM
%along both the dispersion and cross-dispersion axes; (iii) extraction
%of the spectra using the traced positions and widths, assuming a
%Gaussian profile for each fiber projected along the cross-dispersion
%direction, while simultaneously correcting for stray light; (iv)
%wavelength calibration and linear resampling of the extracted spectra;
%(v) differential correction for fiber-to-fiber transmission
%variations; (vi) flux calibration based on stars observed by the
%spectrophotometric telescope concurrently with the science exposures;
%(vii) combination of the spectra from the three channels (b, r, and z)
%for each science fiber into a single continuous spectrum, applying
%inverse-variance weighting in the overlap regions between channels;
%(viii) estimation and subtraction of the night-sky background using
%data from the east and west sky-monitoring telescopes; and (ix)
%computation of the astrometric solution for each observation using
%information obtained from the guiding cameras. Throughout the entire
%reduction process, uncertainties are consistently propagated at every
%stage.

The final product of this reduction process is a FITS file containing
a set of row-stacked spectra corresponding to each LVM pointing, where
every row represents an individual science fiber. Separate extensions
store the flux intensity, the associated estimated uncertainty, and
supplementary information such as the final wavelength solution, the
line-of-sight instrumental velocity dispersion (for each fiber and
wavelength), masks for bad pixels, flags identifying broken or
low-quality fibers, and the estimated sky spectrum. An additional
extension includes the spatial mapping of each science fiber on the
sky (the position table), derived from the astrometric solution,
together with further metadata for each fiber (e.g., additional
quality masks). This FITS file\footnote{\url{https://dr19.sdss.org/sas/dr19/spectro/lvm/redux/1.1.1/0011XX/11111/60191/lvmSFrame-00004297.fits}} constitutes the input for the data
analysis pipeline described in the following sections.

Figure \ref{fig:wise} illustrates the exact location of the pointing
in the sky and the field-of-view (FoV) covered by the LVM science
Integral Field Unit (IFU).  The figure comprises a three-color image
covering a $\sim$0.65\textdegree\ size square area centered on the
Helix Nebula \citep[a similar coverage as the observations discussed
by][]{odell98}.  The image was created by combining the W4 (22$\mu$m,
red), W3 (12$\mu$m, green) and W2 (4.6$\mu$m, blue) band images
obtained by the Wide-field Infrared Survey Explorer
\citep[WISE][]{wrig10}, obtained using the Hierarchical progressive
surveys tool
\citep[HIPS][]{hips}\footnote{\url{https://aladin.cds.unistra.fr/hips/}}.
The LVM science fibers included in the exposure analyzed in this study
are shown on top of this image. The location of broken/low-quality
fibers and the incomplete coverage of the hexagonal IFU FoV is clearly
visible in the figure as empty spaces between fibers. Finally,
we highlight the location of regions previously discussed in the
literature for future discussion \citep[][]{odell98,henry99}.
Fig. \ref{fig:wise} clearly shows that most of the extent of the
{ Helix Nebula} is covered by the actual pointing, however, the coverage
is not 100\%\ complete.  There are low intensity regions that are
clearly outside the footprint of the explored pointing.

%%%%%%%%%%%%%%%%%%%%%%%%%%%%%%%%%%%%%%%%%%%%%%%%%%%%%%%%%%%%%%%%%%%%%%%5
\begin{figure*}
        \centering                
        \includegraphics[width=0.89\textwidth,clip,trim=180 180 10 95]{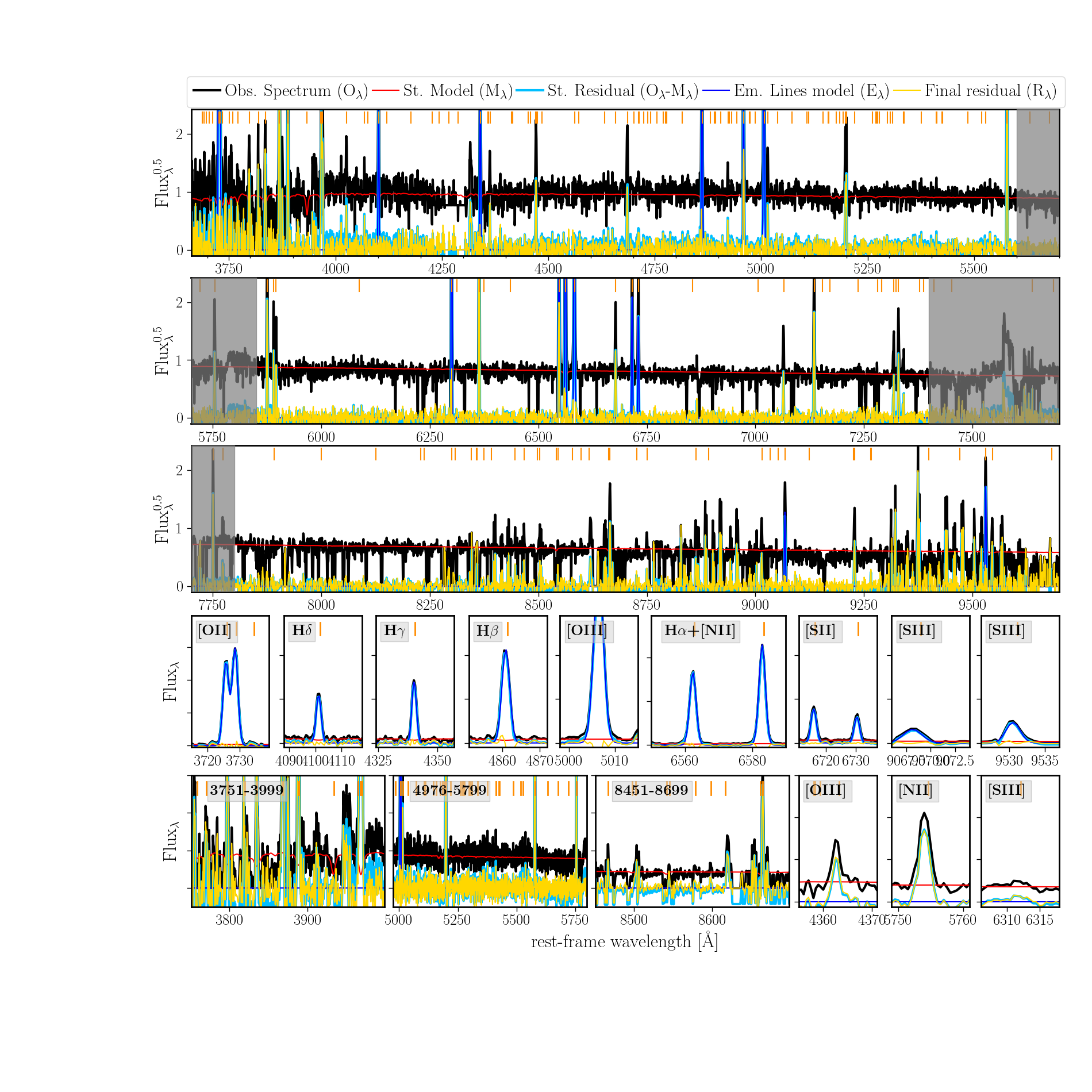}
        \captionof{figure}[]{Integrated spectrum of the { Helix Nebula} across the entire field-of-view of the analyzed LVM pointing (black solid line), together with the results from the LVM-DAP analysis, i.e.,  the best-fitting stellar model (red solid line), the model of the strongest emission lines fitted with Gaussian functions (blue solid line), the residual after subtraction of the stellar model (cyan solid line), and the combination of both models, including a final correction to the residual shape (yellow solid line). The first three panels, from top to bottom, display the wavelength ranges covered by the blue (1st row), green (2nd row), and infrared (3rd row) arms that compose the LVM spectrograph. Shaded areas indicate the overlapping regions between these arms. The insets in the two bottom rows show zoom-ins on selected wavelength intervals. The panels in the 4th row illustrate the quality of the modeling for the strongest emission lines based on the parametric analysis described in the text. Finally, the bottom row of panels presents three wavelength ranges (3751–3999~\AA, 4976–4799~\AA, and 8451–8699~\AA) that highlight specific spectral features discussed in the text. Three additional insets display close-ups around the auroral lines most commonly used to determine the electron temperature in ionized nebula: \oiii~$\lambda4363$, \nii~$\lambda5755$, and \siii~$\lambda6312$. The flux scales in these three insets are identical to facilitate comparison among the lines. In all panels, orange vertical lines indicate the emission lines analyzed using the non-parametric procedure described in the text.}
        \label{fig:spec}
    \end{figure*}
%%%%%%%%%%%%%%%%%%%%%%%%%%%%%%%%%%%%%%%%%%%%%%%%%%%%%%%%%%%%%%%%%%%%%%%5

\section{ANALYSIS}
\label{sec:ana}

We apply the LVM Data Analysis Pipeline (LVM-DAP) to the Helix
nebula pointing described above. The LVM-DAP was presented and
described in detail in \citet{lvmdap}, and it was distributed
publicly \footnote{\url{https://github.com/sdss/lvmdap}}. We present here a brief
summary, highlighting the modifications introduced since it
was first presented. An updated schematic of analysis flow
performed by the LVM-DAP is presented in Figure \ref{fig:scheme}.

The LVM-DAP employs the spectral fitting algorithms implemented in
{\sc pyFIT3D} \citep{pypipe3d}, a Python~3 package designed to
decouple and extract the physical information associated with both the
stellar and ionized gas components in integral-field spectroscopic
(IFS) data. The LVM-DAP follows the general structure of {\sc
  pyPipe3D} \citep{pipe3d,pypipe3d}, with specific adaptations to the
characteristics of the LVM observations, including the spectral
resolution, wavelength coverage, and in particular the nature
of the stellar component sampled by each resolution element (fiber). Each
individual fiber spectrum is analyzed independently, without spatial
binning, ensuring that the spatial information inherent to the LVM’s
ultra-wide IFU is fully preserved.

The first stage of the analysis focuses on characterizing the stellar
continuum. This is achieved by fitting each observed spectrum with a
linear combination of templates from a library of RSP (resolved
stellar populations). In contrast to the usual analysis performed using
other tools such as {\sc pyPipe3D}, {\sc pPXFs} or {\sc STARLIGHT}
\citep[][]{pypipe3d,ppxf,starlight}, which are based on the stellar
synthesis method \citep[][]{bc2003,conroy13}, each template does not
represent the integrated spectrum of a single stellar population (SSP)
with a given age, metallicity, and abundance pattern, formed in a
single burst of star-formation. The varying physical size sampled by
the LVM apertures, depending on the distance within the galaxy of the
captured targets, and its range of values (from a few to a few hundred
of parsecs), does not guarantee a full sampling of the mass
function. Thus, the basic hypotheses behind the stellar synthesis
method are not guaranteed \citep[e.g.][]{cerv13a}. For this reason we introduced the use of RSP, a set of stellar templates, each one
representative of a distribution of physical parameters (Teff, log(g), [Fe/H]
and [$\alpha$/Fe]), characterized by a probability distribution function (PDF),
minimizing and controlling the degeneracies in the space of observed
parameters (i.e., the spectra). The details of how the RSP templates
are defined and created are also explained in \citet{lvmdap} too.
For the current analysis we adopted a set of RSP templates created
based on the MaStar stellar library \citep{mastar}, classified
using the CoSha code \citep{cosha}, which are distributed in the LVM
github repository.

%Each observed spectrum is modeled as the convolution of these templates
%with a line-of-sight velocity distribution (parameterized by the
%stellar velocity $v_*$ and velocity dispersion $\sigma_*$) and
%attenuated by dust with an extinction $A_{V,*}$. These non-linear
%parameters ($v_*$, $\sigma_*$, and $A_{V,*}$)are derived through a
%pseudo-random brute exploration of the space of parameters within a pre-defined
%range of value.
%A limited sub-set of the RSP library, comprising 12 templates,
%is used during this process. The main difference between the procedure
%described in \citet{lvmdap} and the currently adopted one at this stage
%is that for the dust attenuation both the observed spectra and the templates
%are binned and resampled within 50\AA\ to speed-up the process and allow
%its implementation to lower S/N levels.

Two distinct RSP template libraries are employed in the LVM-DAP. The
first one, optimized for efficiency, is used to derive the non-linear
parameters that characterize the stellar kinematics and dust
attenuation ($v_*$, $\sigma_*$, and $A_{V,*}$). This first
library is deliberately small to minimize computational cost and
degeneracies, comprising 12 RSP templates in the current
analysis. During this analysis, which performs a pseudo-random brute-force
exploration of the space of the non-linear parameters within a
pre-defined range of values, the strongest emission lines are masked.

This procedure provides a first/preliminary model of the stellar
spectrum. This stellar model is used to create a continuum-subtracted,
gas-pure spectrum, which is used to create a first model of the
emission lines (following a procedure described below). This emission
line model is then subtracted from the original spectrum to create a
gas-free spectrum.  This spectrum is then fitted using a second, more
extensive RSP library, covering a finer grid of stellar properties and
adopting the non-linear parameters derived in the first step (comprising 108 templates, in the current analysis). In this way, the best stellar model,
the coefficients of the decomposition, and the light-weighted average of the stellar properties were derived (see Fig. \ref{fig:scheme})

The DAP characterizes the emission lines using two complementary
procedures: (i) a \textit{parametric} Gaussian fitting of the
strongest emission lines, providing integrated flux
($F_{\mathrm{int}}$), velocity ($v_{\mathrm{gas}}$), and velocity
dispersion ($\sigma_{\mathrm{gas}}$) for each line system; and (ii) a
\textit{non-parametric} moment-weighted analysis applied to the large
set of emission lines, delivering fluxes, velocities, dispersions, and
equivalent widths (EW). The parametric analysis comprises 15 emission
lines, while the non-parametric analysis currently including 215
emission lines. In the current version of the LVM-DAP (v1.1.0), the
parametric analysis is performed twice \citep[contrary to the
previously published version][]{lvmdap}: the first time after the
derivation of the non-linear parameters of the stellar population, and
the second time after the derivation of the best stellar model using the
full RSP library. The results of both analyses are stored and
delivered separately in the final dataproducts. On the contrary, the
non-parametric analysis is performed only once, using the best stellar
model to create the gas-pure spectrum and adopting as initial guess
values the results of the 1st parametric analysis (see
Fig. \ref{fig:scheme}).

In addition to the double evaluation of the properties of the emission
lines based on the parametric analysis, the current version of the
LVM-DAP has introduced an estimation, for each explored emission line, of
(i) the velocity dispersion in units of km s$^{-1}$, corrected by
the instrumental LSVD provided by the DRP for each fiber and wavelength,
and (ii) the corresponding reduced $\chi^2$. These two values are derived
for both the parametric (2nd evaluation) and non-parametric analyses.
For the former, the $\chi^2$ is derived by just evaluating the final model
using the derived parameters. For the non-parametric method we assume
that the line profile follows a Gaussian function using the values derived
by this procedure. The results are stored in a new table included in the
LVM-DAP dataproducts, together with the adopted instrumental LSVD for
each emission line.

% --------------- DAP file ------------------------%
\begin{table*}[ht]
\RaggedRight
\caption{Description of the DAP file containing the dataproducts of the analysis}
\label{tab:hdu} 
\begin{tabular}{clll}
\toprule
HDU	&  EXTENSION & \# Rows & \# Columns\\
\midrule
  0  & PRIMARY       &  &  \\      
  1  & PT            &  \#spec &  6   \\
  2  & RSP           &  \#spec & 1 + \#RSP \\
  3  & COEFFS        &  \#spec $\times$ \#RSP & 13 \\
  4  & PM\_ELINES    &  \#spec $\times$ \#PM\_EL & 10 \\
  5  & NP\_ELINES\_B &  \#spec &  1+\#NP\_EL\_B $\times$8\\
  6  & NP\_ELINES\_R &  \#spec &  1+\#NP\_EL\_R $\times$8\\
  7  & NP\_ELINES\_I &  \#spec &  1+\#NP\_EL\_I $\times$8\\
  8  & PM\_KEL       &  \#spec  $\times$ \#PM\_EL & 10 \\
  9  & ELINES\_SIGMA\_CHI  & \#spec & 2(2\#PM\_EL+\#NP\_EL)\\
 10  & ELINES\_CHI2\_AVG  & 2\#PM\_EL+\#NP\_EL & 4 \\
 11  & INFO          &  \#param & 2 \\
\bottomrule
\end{tabular}
\tabletext{Structure of extensions included in the delivered DAP file, where: (i) \#spec is the number of science spectra (or fibers) included in the analyzed {\tt Tile} or RSS (row-stacked spectra) frame; (ii) \#RSP is the number of
templates/spectra included in the stellar library; (iii) \#PM\_EL is the number of individual
models (emission lines) included in the parametric analysis of the ionized gas emission lines, and
(iv) \#NP\_EL\_BAND is the number of emission lines included in the non-parametric analysis for each BAND (B, R and I) corresponding to each arm of the spectrograph, and (v) \#NP\_EL is the total number of emission lines analyzed using the non-parametric procedure. 
}
\end{table*}
% --------------- DAP file ------------------------%

Prior to running the full fiber-by-fiber analysis, the DAP
analyzes an integrated spectrum obtained by coadding
all science-fiber spectra of the pointing, weighted by their
inverse variance. First, the average
redshift/velocity within the considered frame is derived by
performing a cross-matching of the observed wavelength of
H$\alpha$ and [\ION{N}{II}] doublet with their rest-frame counterparts. This
derivation of the velocity is used to readjust the range of
explored parameters in any subsequent analysis. Then, the full
analysis described above, following the scheme outlined in
Fig. \ref{fig:scheme}, is applied to this integrated spectrum. The
results of this analysis are used to define reliable initial
guesses and parameter ranges for both the stellar and ionized gas
components for the full analyzed dataset, improving the accuracy
and stability of the individual fits.

The result of this analysis is shown in Figure \ref{fig:spec}. The
integrated spectrum is shown together with the best fit model for
the underlying stellar population and the model for the emission lines
analyzed using the second iteration of the parametric procedure. The
spectrum has been divided into the three wavelength ranges covered by
the three arms of the spectrograph, with a set of insets highlighting
(i) the strongest emission lines within the considered wavelength
range, (ii) three particularly relevant wavelength ranges for the
stellar population analysis (centered on the K+H Ca doublet, \hb\, and the
CaT), and (iii) the wavelength ranges centered on the three most
relevant auroral lines ([\ION{O}{iii}]4363, [\ION{N}{ii}]5755 and
[\ION{S}{iii}]6312).  A visual inspection of this figure clearly
highlights the existing problems with the sky subtraction in the
currently released dataset, reduced using version 1.1.1 of the DRP. The multiple apparent absorptions (and emissions) at any wavelength, but in particular in the regime corresponding to the infrared arm, are a clear consequence of this problem. This is significantly improved in the newer versions of the DRP (above version 1.2.0), currently under development, particularly after introducing the correction for telluric absorptions. This issue is primarily affecting the stellar continuum,
which is very weak and not very suitable for a proper stellar
decomposition.  We will discuss that later on. However, despite
these problems the nebular emission lines analyzed using the parametric
procedure present a very good S/N and accurate modeling.
Furthermore, even some weak features like the auroral lines shown in the maps, have enough signal-to-noise levels to allow the determination of physical parameters of the ionized nebulae (e.g. electron temperature). We should note
here that this distributed dataset corresponds to a frame taken during
the science commissioning of the instrument, when the full observing
procedure (including the sky estimation) adopted along the formal
survey period was still not in place and it was reduced with an
under-development version of the DRP. A spectrum more representative
of the current quality of the LVM data was presented in \citet{lvmdap}, \citet{satt25}, \citet{gonz25}, \citet{villa25} and \citet{hild25}.

Throughout all steps of the analysis, uncertainties are propagated via
Monte Carlo simulations that perturb the input fluxes according to the
errors provided by the DRP. Thus, for each of the physical parameters delivered by our analysis, there is an associated error estimation. How these errors are representative
of real uncertainties was discussed in \citet{lvmdap}.

\subsection{Night Sky 2nd Order Correction}
\label{sec:sky}

As indicated before, the analyzed dataframe presents some defects due
to an imperfect sky subtraction (e.g., Fig. \ref{fig:spec}). Our current
understanding of this problem is that it arises from a
combination of a problem in the estimation of the night-sky spectrum
itself and an imperfect estimation of the fiber-to-fiber transmission
correction (Mej\'\i a-Narv\'aez et al., in prep.). We acknowledge this
problem in the data reduction applied to the publicly accessible frame
which is indeed properly addressed in under-development versions of the
DRP. For the current dataset we present an {\it a posteriori}
correction for those emission lines that are clearly affected by this
issue, that we present here.

First, we select two regions within the FoV of the IFU
(Dec$>$-20.57\textdegree\ or Dec$<$-21.0\textdegree\ ), corresponding either to 
areas of low-intensity (for the strong emission lines) or those that should
correspond directly to night-sky emission. For each emission line we determined the median flux intensity ($med_{f-back}$) within these
two regions along with the corresponding standard deviation
($\sigma_{f-back}$)\footnotetext{{ We note that different approches, including the an average and a set of weighted-average was explored before deciding to adopt this final approach}}. For those emission lines in which (i) the median flux is above a certain threshold in units of the standard deviation ($n_{\sigma}$; this means there is non-negligible signal) and (ii) the standard deviation is below an additional threshold, $lim_{\sigma}$ (in order to avoid structure in the detected signal, separating it from a homogeneous background contribution), we subtract $med_{f-back}$ from the flux intensity of the considered emission line for all fibers. The
$n_{\sigma}$ and $lim_{\sigma}$ values (1.5 and 10$^{-13}$
erg\,s$^{-1}$\,cm$^{-2}$) were the same for all emission lines, being
selected in an iterative process in which $med_{f-back}$ of the
corrected lines is minimized.

{ The described correction was derived for all emission lines. However, for only four of the 215 analyzed lines does it produce a non-negligible effect (i.e., a modification above the noise level):}
[\ION{N}{II}]5755, [\ION{O}{I}]6300,
[\ION{O}{I}]6364, and HI8665. These lines are among the most relevant (and well explored in previous literature) to explore the ionization structure and the physical properties of the nebula and its physical properties, what justifies the described
correction. Furthermore, it is worth noticing that in the case of \ION{N}{II}5755, the background bias is not homogeneous across the FoV, showing a pattern clearly associated to the six fiber-bundles that
configure the LVM science IFU \citep[][]{herbst24}. For this
particular case the $med_{f-back}$ is subtracted only in those regions
in which a clear background excess intensity is evident.

In Appendix \ref{app:cor} we show an example of the effect of the
described correction on a particular emission line.

\subsection{Emission lines Golden Sample}
\label{sec:good}

Once this correction is performed we explore which lines among the 215
analyzed ones are detected above the background noise.  For doing so
we select two circular apertures of 3$\arcmin$ located at (i) the
central region of the nebula (-20.775\textdegree\ , 337.450\textdegree), and
(ii) a region corresponding to the brightest area of the nebular ring,
(-20.850\textdegree\ , 337.400\textdegree). These two areas are selected
acknowledging the different spatial distribution expected for
different emission lines arising from different ionizing regions
within the nebula. We then obtain the median flux intensity in both
apertures ($med_{f-neb}$), together with the integrated flux intensity
and its corresponding error across the entire FoV. We mark as good
detections those emission lines in which (i)
$med_{f-neb}>$1.5$\sigma_{f-back}$ for any of the considered regions,
and (ii) the integrated flux intensity is at least three times its
error ($\approx$3$\sigma$ detection). These dual criteria are
selected to guarantee not only a sufficient S/N in the integrated
quantities, but that the detected flux corresponds to areas actually
covered by the nebula. A total of 54 emission lines fulfill both
criteria (i.e., $\sim$26\% of the original analyzed set of emission
lines). We define this { sub-set} as our emission lines golden sample.

\section{RESULTS}
\label{sec:res}

% Lines affected by blending or high Chi-SQ
% HI 3797.9
%HeI 3819.61
%HI 3835.38
%HeI 3888.65
%HI 3889.05
%HeI 3964.73
%$[$\ION{NeIII}{}$]$ 3967.46
%CaII 3968.47
%H$\epsilon$ 3970.07

%----------------------------------------EL HELIX NP-----------------------
\begin{table*}[ht]
\RaggedRight
\caption{Integrated flux intensities of the Helix Nebula emission line golden sample.}
\label{tab:fe}
\begin{tabular}{lcc|lcc|lcc}
\toprule  
Name & $\lambda$$^1$ & flux$^2$ & name & $\lambda$$^1$ & flux$^2$ & name & $\lambda$$^1$ & flux$^2$  \\
%     & (\AA) & (10$^{-13}$ erg/s/cm$^2$) &  & (\AA) & (10$^{-13}$ erg/s/cm$^2$) & & (\AA) & (10$^{-13}$ erg/s/cm$^2$)\\
\midrule
$[$\ION{OII}{}$]$  &  3726.03 & 4693.4  $\pm$  4.7 & $[$\ION{FeII}{}$]$  & 4474.91 & 99.1  $\pm$ 18.3 & HeI  & 6678.15 & 128.4  $\pm$  3.4 \\
$[$\ION{OII}{}$]$  &  3728.82 & 6142.6  $\pm$  4.4 & HeII  & 4685.68 & 248.1  $\pm$  3.5 & $[$\ION{SII}{}$]$  & 6716.44 & 403.2  $\pm$  4.2 \\
HI$^{3}$  &  3797.90 & 228.2  $\pm$  7.7 & $[$\ION{ArIV}{}$]$  & 4711.33 & 27.9  $\pm$  2.6 & $[$\ION{SII}{}$]$  & 6730.82 & 290.0  $\pm$  1.8 \\
HeI$^{3}$  &  3819.61 & 154.7  $\pm$ 29.2 & HeI  & 4713.14 & 18.8  $\pm$  2.8 & HeI  & 7065.19 & 85.3  $\pm$  2.0 \\
HI$^{3}$  &  3835.38 & 249.7  $\pm$ 19.9 & H$\beta$  & 4861.36 & 2382.1  $\pm$  7.8 & $[$\ION{ArIII}{}$]$  & 7135.80 & 560.9  $\pm$  1.5 \\
$[$\ION{NeIII}{}$]$  &  3868.75 & 2100.8  $\pm$  6.0 & $[$\ION{OIII}{}$]$  & 4958.91 & 3690.3  $\pm$ 83.9 & $[$\ION{OII}{}$]$  & 7318.92 & 131.6  $\pm$  3.4 \\
HeI$^{3}$  &  3888.65 & 637.4  $\pm$  3.4 & $[$\ION{OIII}{}$]$  & 5006.84 & 11820.9  $\pm$ 25.5 & $[$\ION{CaII}{}$]$  & 7323.88 & 75.0  $\pm$  2.9 \\
HI$^{3}$  &  3889.05 & 613.4  $\pm$  3.4 & $[$\ION{NI}{}$]$  & 5197.90 & 126.6  $\pm$  2.0 & $[$\ION{OII}{}$]$  & 7329.66 & 116.0  $\pm$  2.4 \\
HeI$^{3}$  &  3964.73 & 72.1  $\pm$  3.5 & $[$\ION{NI}{}$]$  & 5200.26 & 146.7  $\pm$  2.0 & $[$\ION{ArIII}{}$]$  & 7751.06 & 163.7  $\pm$  4.7 \\
$[$\ION{NeIII}{}$]$$^{3}$  &  3967.46 & 752.4  $\pm$  2.9 & HeII  & 5411.52 & 15.9  $\pm$  4.6 & HeII  & 8236.77 & 11.3  $\pm$  1.6 \\
CaII$^{3}$  &  3968.47 & 633.9  $\pm$  3.1 & $[$\ION{FeII}{}$]$  & 5412.64 & 14.8  $\pm$  4.6 & HI  & 8239.24 & 6.9  $\pm$  1.7 \\
H$\epsilon$$^{3}$  &  3970.07 & 463.4  $\pm$  3.0 & $[$\ION{NII}{}$]$  & 5754.59 & 120.2  $\pm$  3.2 & HI  & 8598.39 & 14.0  $\pm$  2.0 \\
HeI  &  4026.19 & 78.0  $\pm$ 10.2 & HeI  & 5876.00 & 460.2  $\pm$  2.6 & $[$\ION{CI}{}$]$  & 8727.13 & 37.1  $\pm$  4.7 \\
H$\delta$  &  4101.77 & 642.9  $\pm$  4.6 & $[$\ION{OI}{}$]$  & 6300.30 & 854.6  $\pm$  7.0 & HI  & 8750.47 & 59.0  $\pm$  2.0 \\
H$\gamma$  &  4340.49 & 1067.1  $\pm$  5.4 & $[$\ION{OI}{}$]$  & 6363.78 & 283.2  $\pm$ 15.5 & HI  & 8862.78 & 59.8  $\pm$  4.1 \\
$[$\ION{OIII}{}$]$  &  4363.21 & 51.9  $\pm$  7.4 & $[$\ION{NII}{}$]$  & 6548.05 & 3717.6  $\pm$ 68.8 & $[$\ION{SIII}{}$]$  & 9069.00 & 169.1  $\pm$  3.2 \\
HeI  &  4471.48 & 144.9  $\pm$  7.6 & $[$\ION{NII}{}$]$  & 6583.45 & 11206.8  $\pm$ 22.8 & HI  & 9229.02 & 62.9  $\pm$  0.7 \\
\bottomrule
\end{tabular}    
\tabletext{(1) Wavelength is expressed in \AA; (2) flux intensity is expressed in 10$^{-13}$ erg/s/cm$^2$. All fluxes are based on the LVM-DAP non-parametric analysis; (3) Emission-lines 
which a $\chi^2$ three times larger than the average, which flux derived using the non-parametric procedure has to be taken with care, as it could be blended with other nearby emission lines or night-sky residuals.}
\end{table*}
%----------------------------------------------------------------

\subsection{DAP dataproducts}
\label{sec:res_dp}

The main result of this exploration is the set of dataproducts derived
by LVM-DAP for the analyzed exposure. These dataproducts are delivered
as a multi-extension FITS-file (the DAP
file\footnote{\url{https://ifs.astroscu.unam.mx/LVM_DR19_Helix/Helix_DR19_new.dap.fits.gz}}),
comprising a set of FITS-tables in which there are stored the
different parameters (and errors) for each science fiber, and some
additional extensions comprising relevant information about the
analyzed pointing and the parameters adopted during the analysis.

The format was already described in \citet{lvmdap}. For the currently
delivered data product, each extension comprises (i) the
header of the original dataframe for traceability ({\tt PRIMARY}), and
a set of FITS tables including (ii) the mapping or position table of
each science fiber on the sky ({\tt PT}), (iii) the luminosity
weighted average values of the physical properties of the stellar
component derived by the stellar decomposition process ({\tt RSP}),
(iv) the actual weights obtained in this decomposition required to
reconstruct the stellar spectral model and generate the PDF of the
stellar properties ({\tt COEFFS}, and the properties of the emission lines
derived by (v) the 1st analysis using the parametric fitting process
({\tt PM\_ELINES}), (vi) the non-parametric procedure, separated in each
spectrograph arm ({\tt NP\_ELINES\_B}, {\tt NP\_ELINES\_R} and {\tt
  NP\_ELINES\_I}, (vii) the 2nd analysis using the parametric
procedure ({\tt PM\_KEL}), (viii) the velocity dispersion corrected by
instrumental resolution in km~s$^{-1}$ and the reduced $\chi^2$ of the
fitting to each emission line ({\tt ELINES\_SIGMA\_CHI}), (ix) the
actual instrumental resolution correction applied and the average
reduced $\chi^{2}$ for each analyzed emission line ({\tt
  ELINES\_CHI2\_AVG}), and finally (x) a table including all the
configuration parameters adopted to run the LVM-DAP in the current
analysis ({\tt INFO}).

Table \ref{tab:hdu} summarizes the format of the delivered DAP file,
describing the size of each extension. It is important to note that
all tables included in each extension containing dataproducts (all but
{\tt PRIMARY} and {\tt INFO}) contain a column with a unique key
({\tt id}) built using the exposure number of the observed frame ({\tt
  EXPNUM}, 4297 in this particular case), and the identification
number of the analyzed fiber in the original {\tt SLITMAP} extension
provided by the DRP ({\tt FIBERID}). This way,
$\rm id = EXPNUM.FIBERID $, is a unique identification key for a
particular fiber corresponding to a particular location in the sky
observed by a particular LVM exposure. Using this {\tt id} it is very
easy to either combine the content of the different tables, using it
as a key parameter to join columns, and/or combine the dataproducts
derived from different exposures \citep[for instance, to create
mosaics, as those shown in Fig. 8-11 of ][or combine dithered
observations]{drory25}.

In addition to the standard DAP file, we deliver the result of the
background correction described in
Sec. \ref{sec:sky}\footnote{\url{https://ifs.astroscu.unam.mx/LVM_DR19_Helix/Helix_DR19_cor.fits.gz}}.
Contrary to the previous file this comprises a single FITs table in
which the unique {\tt id} described before is included, along with the {\tt RA} and
{\tt Dec} describing the position of the fiber in the sky, and the corrected
flux intensities for the considered emission lines. We should stress
that this is not a standard LVM-DAP product as the implemented {\it a
  posteriori} correction would not be required when the updated
version of the DRP is implemented.

%To facilitate the use of the delivered DAP dataproducts we have included
%a set of tools in the LVM-DAP that will be described in 

\subsection{Integrated emission line fluxes}
\label{sec:res_int}

As indicated in Sec. \ref{sec:ana} the analysis that comprises the LVM-DAP
is first applied to an inverse-variance weighted integrated spectrum
(see Fig. \ref{fig:spec}). This procedure provides same set
of parameters included in the DAP-file described in
Sec. \ref{sec:res_dp} for this integrated spectrum. However, due to
the weighting procedure those values
may not be directly adopted as a good representation of the real
integrated properties of the considered exposure, in this particular
case of the { Helix Nebula}. A better set of values is obtained by coadding the
flux intensities included in the different emission line extensions
included in the DAP file (see Table \ref{tab:hdu}).

Table \ref{tab:fe} lists the integrated flux intensities derived from
this analysis. There are few studies that report the absolute
integrated fluxes for this nebula, and certainly none that provides
measurements for the complete list included here. Early
observations using narrow-band imaging reported an integrated flux for
[\ION{O}{III}] and H$\beta$ of
$F(\mathrm{[O\,III]}\,\lambda5007)=19.4\times10^{-10}$ and
$F(\mathrm{H}\beta)=3.37\times10^{-10}\,\mathrm{erg\,cm^{-2}\,s^{-1}}$
\citep{odell98}, superseding an earlier measurement of
$F(\mathrm{H}\beta)=4.5\times10^{-10}\,\mathrm{erg\,cm^{-2}\,s^{-1}}$
by \citet{odell62}. For these two emission lines, our LVM integrated
spectrum yields $F(\mathrm{[O\,III]}\,\lambda5007)=11.82\times10^{-10}\,\mathrm{erg\,cm^{-2}\,s^{-1}}$ and
$F(\mathrm{H}\beta) = 2.38
\times10^{-10}\,\mathrm{erg\,cm^{-2}\,s^{-1}}$, respectively.  Thus,
the LVM absolute fluxes are lower than the historical full-nebula
values by $\sim\!39\%$ for [O\,III]\,5007 and $\sim\!29-47\%$ for
H$\beta$. On the other hand, the line ratio derived using the LVM
values, $F(\mathrm{[O\,III]}\,\lambda5007)/F(\mathrm{H}\beta)=4.96$ is
within the two historical values, 4.3 and 5.8, differing by only
$\sim 15\%$. The LVM-IFU FoV, with a hexagonal area of
$\sim$0.45\textdegree$^2$, covers most of the optical extent of the
nebula, and certainly the brightest regions. However, it is still
smaller than the area covered by the discussed narrow-band images: a
rectangular FoV of $\sim$0.65\textdegree\ size \citep{odell98},
covering an area of $\sim$1.17\textdegree$^2$ (i.e., $\sim$2.6
times larger). Besides that, the filling factor of the LVM-IFU, considering
fibers of 35.3$\arcsec$, is $\sim$83\%. The combined effect of a
smaller FoV and the non-complete filling factor may well explain this
discrepancy.

A rough estimation of the flux lost due to these effects could be
obtained by comparing the integrated flux of the WISE W3-band image
included in Fig. \ref{fig:wise} with the one observed through the LVM
science fibers. The emission at $\sim$12$\mu$m traces the H$\alpha$
luminosity, following a correlation that it is frequently used to
explore the integrated SFR in galaxies
\citep[e.g.][]{leroy21,colombo25}.  This correlation holds
irrespective of the photoionizing source, as the physical nature of
the connection between both emissions is the same: the ultraviolet
emission that produces the ionization is the one that it is reprocessed
and emitted at FIR due to the dust. Based on this analysis we estimate
$\sim$31\%\ of lost flux, which is reasonably in agreement with the
reported discrepancy.

The values reported in Tab. \ref{tab:fe} correspond to the non-parametric
analysis, a procedure that it is not optimized to handle the possible contamination
by nearby/blended emission lines or night-sky residuals. We have labeled all those lines that could
be affected by this effect in Tab. \ref{tab:fe}, based on the $\chi^2$ of the comparison of the
gas-pure spectra with a Gaussian model created using the parameters derived by the non-parametric procedure.
The effect may affect more clearly the weak emission lines located nearby much stronger ones (e.g., CaII3968 instead of \he). 
The reported fluxes for those lines should be taken with care.

%Thus, the values reported in Tab. \ref{tab:fe},
%although they are somehow representative of the integrated values
%in the { Helix Nebula} they may be affected by an aperture effect.
%{ TBW, check with GALEX and WISE!!!!}

\subsection{Spatial distribution of the ionized gas}
\label{sec:res_map}    

%%%%%%%%%%%%%%%%%%%%%%%%%%%%%%%%%%%%%%%%%%%%%%%%%%%%%%%%%%%%%%%%%%%%%%%5
\begin{figure*}
  \centering
  \includegraphics[width=0.89\textwidth,clip,trim=30 120 0 200]{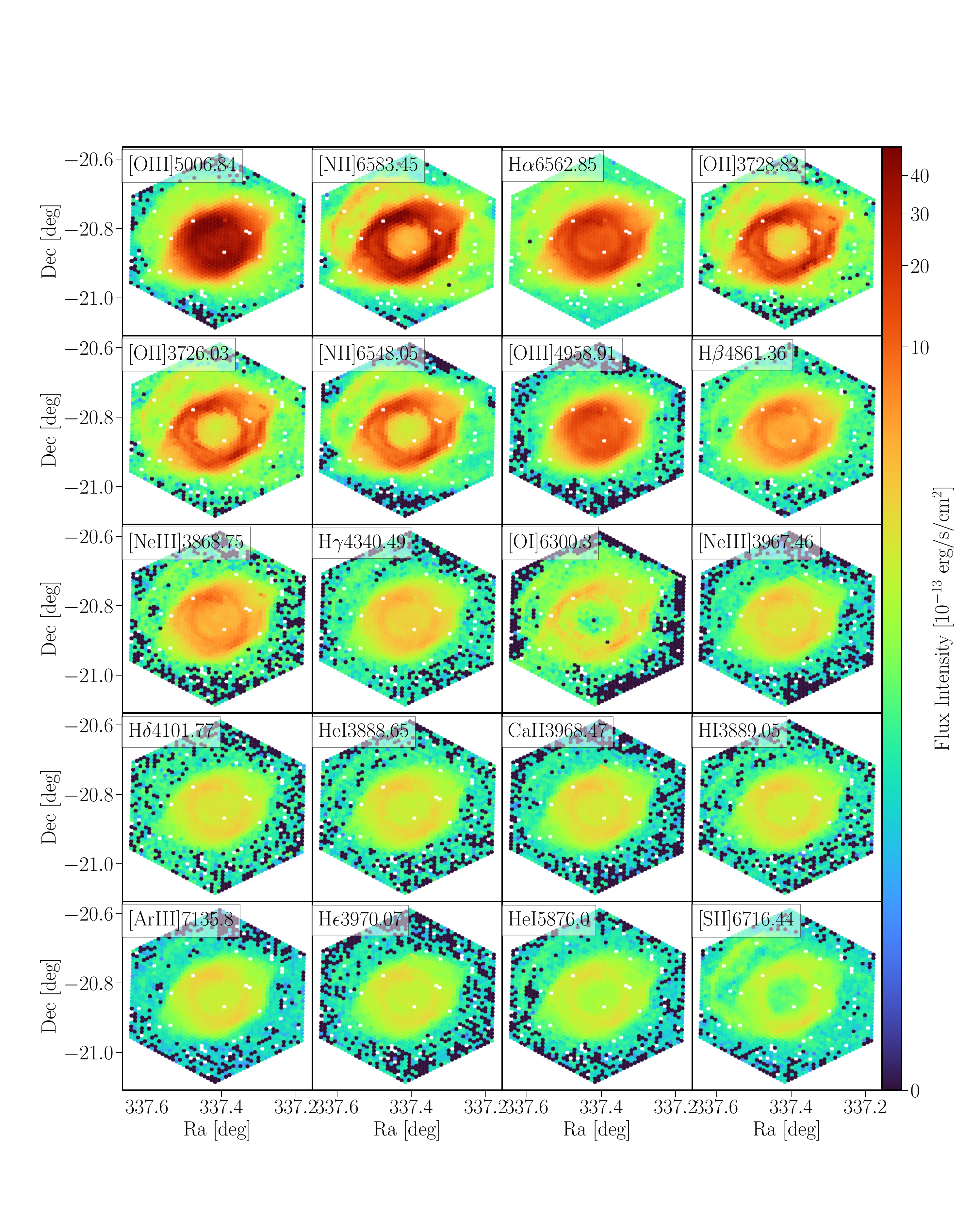}  
   \captionof{figure}[]{Example of the analysis performed by LVM-DAP to recover the properties of the ionized gas emission lines. Each panel shows
 the distribution across the FoV of the LVM IFU of the flux intensity estimated by the weighted-moment procedure for the 20 brightest emission lines shown in the integrated spectrum of the { Helix Nebula}(Fig. \ref{fig:spec}). The emission lines are ordered from the brightest (top-left) to the faintest (bottom-right). The legend in each panel indicates the represented emission line. The remaining detected emission lines are shown in Appendix \ref{app:fe}}
        \label{fig:fe}
    \end{figure*}
%%%%%%%%%%%%%%%%%%%%%%%%%%%%%%%%%%%%%%%%%%%%%%%%%%%%%%%%%%%%%%%%%%%%%%% 5

%%%%%

Figure \ref{fig:fe} shows the flux intensity maps of the 20
brightest emission lines included in the {\it golden sample}
defined in Sec. \ref{sec:res_int} and listed in Tab. \ref{tab:fe}
(for the remaining lines, the maps are included in Appendix
\ref{app:fe}). The observed spatial distributions reveal the
complex ionization structure of the Helix Nebula, highlighting the
characteristic morphology produced by the interaction between the
ionizing radiation from the central star and the expanding nebular
gas. The overall appearance of the nebula is dominated by a bright
annular structure encircling a lower-surface-brightness core,
surrounded by fainter, more diffuse emission that extends into the
outer halo. The different emission lines display distinct spatial
morphologies, reflecting the stratification of ionization states
within the nebula.

The maps of different Balmer lines (e.g. \Ha\, \Hb, \Hg\ and \He),
which trace the distribution of the ionized hydrogen, effectively delineate the
full extent of the ionized gas. Their relatively smooth brightness
across the main ring indicates that the Balmer emission primarily
arises in regions of moderate ionization, where hydrogen is almost
fully ionized but helium is only singly ionized. The strong contrast
between the bright inner ring and the fainter outer halo suggests that
the density of the ionized gas peaks within this structure, which
likely represents the intersection of a moderately inclined ionization
front with the plane of the sky.

The [\ION{O}{iii}]5007 emission, the brightest emission line among those
observed in the LVM dataset, displays a slightly concentrated distribution toward the inner portions of
the main ring. This behavior reflects the dependence of the O$^{++}$
ion on the availability of photons with energies above 35~eV \citep{oster06}, which
are abundant near the central star \citep{oster06,odell98,meaburn05,odell07}. The resulting [\ION{O}{iii}]
morphology thus traces zones of high excitation, corresponding to gas
fully ionized in hydrogen and helium but not yet depleted of O$^{++}$
through further ionization. The enhanced [\ION{O}{iii}] emission
observed in the inner parts of the ring likely marks the transition
between the He$^{++}$ and He$^{+}$ regions, where the local electron
temperature reaches its maximum as a result of efficient photoelectric
heating and reduced cooling efficiency. Similar distributions are observed
in other high excitation emission lines, like [\ION{Ar}{III}]7136,
[\ION{S}{III}]9069, or [\ION{S}{III}]9531.

In contrast, emission lines like [\ION{N}{ii}]6583, [\ION{O}{ii}]3727,
and [\ION{S}{ii}]6717,31, highlight the low-ionization structures
that lie at and beyond the main ring \citep[e.g.][]{meaburn05}. For these emission lines the
brightest regions are found along the outer boundary of the
[\ION{O}{iii}] bright ring, where the ionization front is tangential
to the line of sight. Nitrogen, oxygen and sulfur are singly ionized
in these regions (e.g., N$^{+}$), coexisting, most probably, with
neutral helium and partially ionized hydrogen. The enhancement of
[\ION{N}{ii}] emission along filamentary features and knot-like
structures, described by \citet{odell98}, suggests the presence of
dense condensations that are externally photoionized and shielded from
the most energetic photons, leading to the observed sharp transition
between high- and low-ionization zones.

%%%%%%%%%%%%%%%%%%%%%%%%%%%%%%%%%%%%%%%%%%%%%%%%%%%%%%%%%%%%%%%%%%%%%%%5
\begin{figure*}
  \centering
  \includegraphics[width=8.5cm,clip,trim=0 0 0 0]{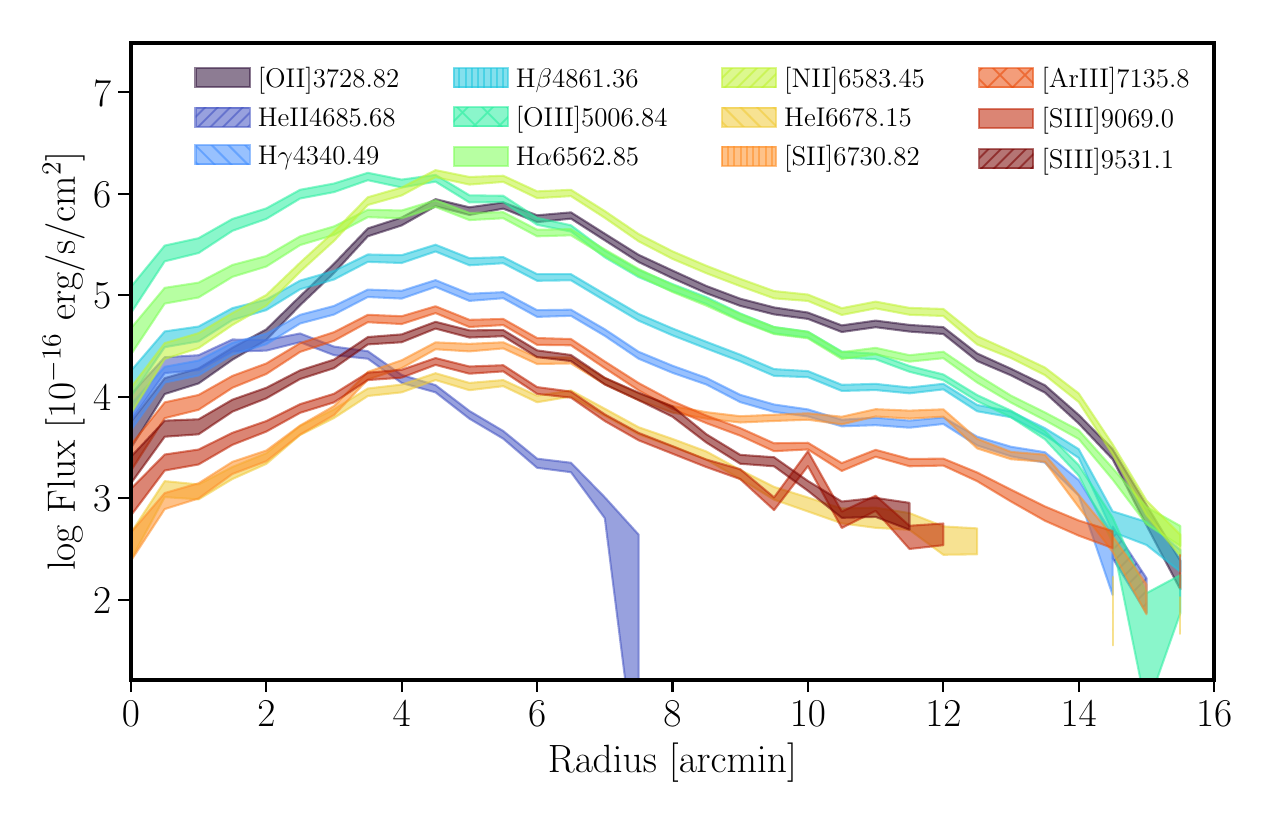}
  \includegraphics[width=8.5cm,clip,trim=0 0 0 0]{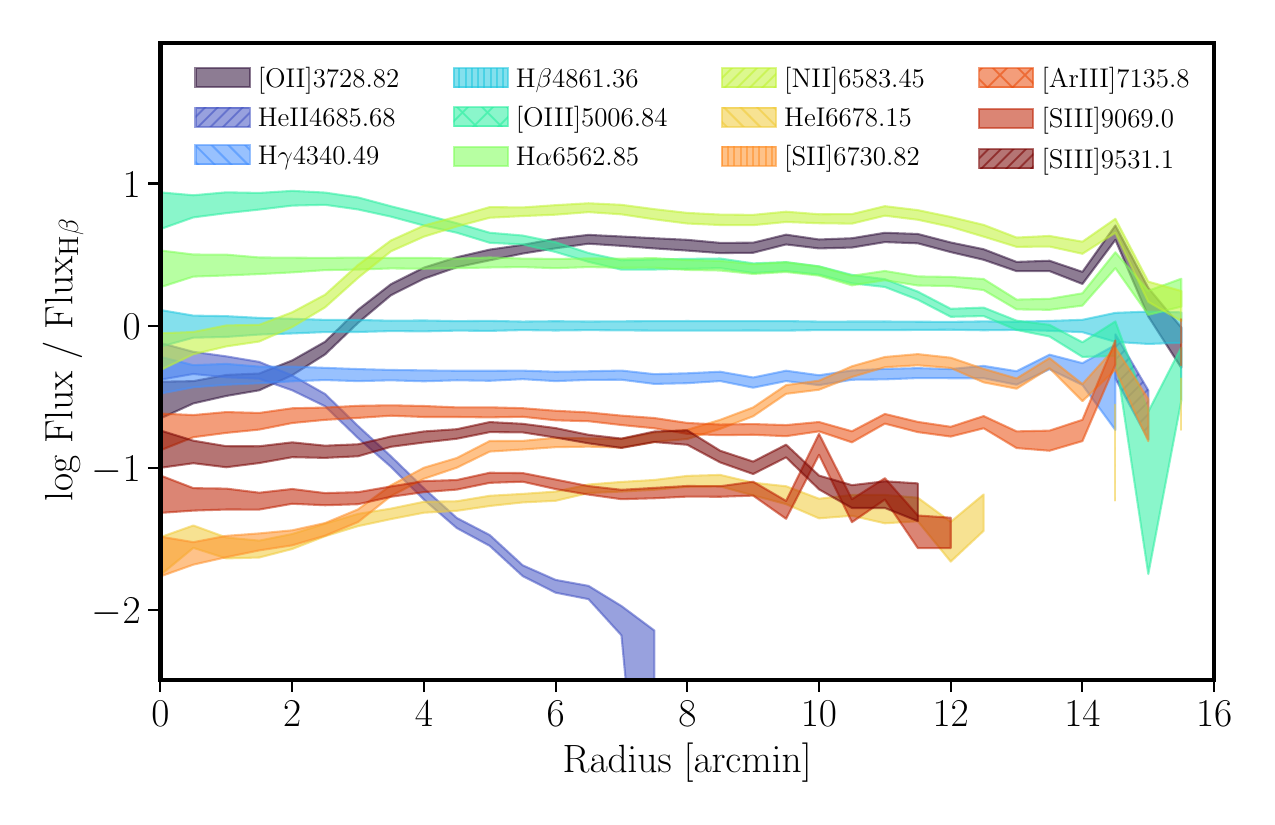}      
  \captionof{figure}[]{Radial profiles of the azimuthally averaged flux intensities in each fiber in absolute values (left-panel), and relative to H$\beta$ (right-panel), for a sub-set of the emission lines analyzed along this study. The filled region in each profile corresponds to the standard deviation across radial bin of 0.5$\arcmin$. Each emission line is represented with a different color and hash-style.  The sub-set comprises both prominent hydrogen recombination lines (H$\alpha$, H$\beta$, H$\gamma$), and key collisional excited transitions ($[$\ION{O}{ii}$]$3737, $[$\ION{O}{iii}$]$5007, $[$\ION{N}{ii}$]$6584, $[$\ION{S}{II}$]$6731, $[$\ION{Ar}{iii}$]$ and $[$\ION{S}{III}$]$9069,9531) that trace the ionization structure of the nebula. The profiles reveal the characteristic bright main ring around $\sim$5$\arcmin$–7$\arcmin$ and the decline toward the outer halo, highlighting differences in excitation between high- and low-ionization species. Similar structure is seen in Fig. \ref{fig:fe}.}
        \label{fig:rad}
    \end{figure*}
%%%%%%%%%%%%%%%%%%%%%%%%%%%%%%%%%%%%%%%%%%%%%%%%%%%%%%%%%%%%%%%%%%%%%%%5

The HeII4686 line outlines a compact, centrally
concentrated region surrounding the hot central star. Similar patterns are found for the other HeII lines included among the golden sample (HeII5411 and HeII8237), despite being considerably weaker. This morphology
indicates that the emission originates within the He$^{++}$ zone,
where helium is fully ionized by photons with energies exceeding
54.4~eV \citep[e.g.][]{oster06}. The circular symmetry of the HeII distribution, and
its confinement well inside the [\ION{O}{iii}] ring, support the scenario
in which the ionizing continuum is dominated by a very hot central star whose hard
radiation field maintains a small but bright He$^{++}$ volume in the
inner nebula \citep{odell98,odell07,henry99}. The relative faintness of this feature compared to
other emission lines, like [\ION{O}{iii}], \Ha\ or \Hb\ is consistent with the limited solid angle
subtended by this high-excitation core \citep{grue92,tyle03}.

Finally, the HeI6678 emission line delineates the region of
intermediate ionization between the compact He$^{++}$ core and the
low-ionization periphery. This line arises from the recombination of
singly ionized helium and traces zones where hydrogen is fully ionized
but helium remains only partially ionized \citep[e.g.][]{oster06}. Its
morphology closely follows that of Balmer lines, but it peaks slightly
inward, indicating the spatial extent of the He$^{+}$ zone produced by
photons with energies between 24.6 and 54.4~eV
\citep[e.g.][]{peim83,benj99}. The smooth distribution of HeI emission
across the main ring suggests that most of the nebular volume is
maintained in this intermediate ionization state, consistent with a
central star that provides a hard, yet not extreme, ultraviolet
radiation field. The relative strength of HeI6678 further supports a
He$^{+}$/H$^{+}$ abundance ratio of about 0.1, typical of a nebula
where helium is nearly fully ionized throughout the main body
\citep{benj99,peim17}. Indeed, this was the helium abundance reported
for the { Helix Nebula} in previous studies \citep{king94}.

In addition to the dominant bright annular structure discussed above, the emission-line maps in Fig.~\ref{fig:fe} reveal two secondary morphological components. First, a fainter but clearly defined inner, ring-like feature visible in some of the strongest emission lines, including \Ha, \Hb, and [\ION{O}{iii}]5007. This structure delineates the boundary of the central cavity surrounding the hot central star and is interpreted as the inner edge of the ionized shell, where the line of sight becomes tangent to the cavity wall. A second, distinct feature is the low-ionization "arm" located in the north-western quadrant of the nebula, which stands out prominently in [\ION{N}{ii}]6583, [\ION{O}{ii}]3729, and the [\ION{S}{ii}]6716,6731 doublet. This structure traces material at larger radii than the main [\ION{O}{iii}] ring and is consistent with externally illuminated, lower-excitation gas associated with the outer ejected layers of the nebula. Its enhanced low-ionization emission suggests higher local densities and partial shielding from the hardest ionizing photons, conditions that favor the survival of N$^{+}$, O$^{+}$, and S$^{+}$ ions. Both structures have been reported and discussed in earlier studies of the Helix \citep[e.g.][]{odell98,henry99,meaburn05}. 

In summary, the observed distribution of the discussed emission lines
demonstrates that the spatial morphology of the Helix
Nebula arises naturally from its three-dimensional ionization
structure \citep{2005RMxAC..23....5O}. The emission-line stratification reflects both the spectral
hardness of the central star and the projection of the ionization
front through a geometrically thick, inclined disk-like shell. These
features are consistent with a photoionization-dominated nebula where
variations in local density and geometry modulate the observed
brightness distribution across different ionic species, as already discussed
in previous studies \citep[e.g.][]{odell98,henry99,odell07}.

\subsection{Radial distribution of emission line fluxes}
\label{sec:res_rad}
    
In previous sections we have determined that the nebula presents,
to first order, a spherical ionization structure, with a central
ionizing source (the PN hot star).  Therefore, it should be well
characterized by the radial distribution of the observed emission
lines. Figure~\ref{fig:rad} (left panel) shows the azimuthally
averaged radial profiles of the flux intensities for a representative subset of these emission lines. 
Those profiles were obtained by deriving the mean and standard deviation of the fluxes within a set of circular annuli of 0.5$\arcmin$ width centered at the location of the central white dwarf that ionizes the nebula. For all lines (and line ratios) it is evident that there is a very small dispersion around the mean value at any radial distance. This feature, per se, strengthens the proposed scenario in which the ionization structure is well represented by a spherical, highly symmetrical, distribution.

The radial behavior of the Balmer lines (\Ha, \Hb, and \Hg) closely follows the overall
surface brightness of the nebula, with a broad maximum corresponding to the
main ring and a smooth decline toward both the inner cavity and the outer halo.
This morphology confirms that the hydrogen recombination emission traces the full
extent of the ionized gas, as discussed in Sec.~\ref{sec:res_map}. The gradual
variation of their intensities with radius indicates that the ionized hydrogen remains
nearly homogeneous within the main ring, consistent with a photoionization-dominated
structure in which density variations dominate over temperature gradients
\citep[e.g.,][]{odell98,henry99,meaburn05,odell07}.

High-ionization species such as [\ION{O}{iii}]5007 or
[\ION{S}{iii}]9069,9531 exhibit more peaked profiles toward the inner
portions of the nebula (with [\ION{O}{iii}]5007 being the most centrally peaked emission line). 
Their maxima occur at smaller radii than
those of the Balmer lines, confirming the strong excitation gradient
inferred from the two-dimensional maps. This concentration reflects
the dependence of the O$^{++}$ and S$^{++}$ ions on photons with
energies above $\sim$35~eV (assuming a constant gas density), which are abundant only close to the
central star \citep{oster06,odell98}. The smooth decline of these
lines outward suggests a gradual softening of the radiation field with
increasing distance, as the high-energy photons are absorbed by the
inner gas layers \citep[e.g.][]{melle95,scho05}.

In contrast, low-ionization lines such as [\ION{N}{ii}]6583,
[\ION{O}{ii}]3729, and [\ION{S}{ii}]6731 peak at larger radii,
delineating the ionization front at the outer boundary of the main
ring (consistent with the distribution shown by the Balmer
lines). Their enhanced emission coincides with the regions where the
ionization front becomes tangential to the line of sight \citep[e.g.][]{odell98}, as seen in
the spatial maps (Fig.\ref{fig:fe}). These lines trace ionized zones with a low degree of ionization, where N, O, and S are singly ionized and coexist to a large extent with neutral helium, as already described in the imaging analyses of \citet{odell98} and the spectroscopic studies by \citet{meaburn05,odell07}.
The contrast between the inner
[\ION{O}{iii}] maximum and the outer [\ION{N}{ii}]/[\ION{S}{ii}]
enhancement reflects the classical ionization stratification expected
in a photoionized shell illuminated by a hot central star \citep[e.g.][]{odell07}.

%These lines trace partially
%ionized zones where N, O, and S are singly ionized, coexisting with
%neutral helium and residual atomic hydrogen, as already described in
%the imaging analyses of \citet{odell98} and the spectroscopic studies
%by \citet{meaburn05,odell07}. 

%Intermediate-ionization species such as HeI6678 display radial
%profiles that peak between those of HeII4686 and the Balmer
%lines, consistent with emission from the He$^{+}$ zone that fills most of the nebular
%volume \citep[e.g.,][]{peim83,benj99,peim17}.

The HeII4686 line shows a compact distribution with a sharp central maximum,
confirming its origin within the innermost He$^{++}$ zone
\citep{odell98,henry99,odell07}. On the other hand, intermediate-ionization species such as HeI6678 display radial profiles that peak well beyond those of the HeII4686 and the Balmer
lines, consistent with emission from the He$^{+}$ zone that fills most of the nebular
volume \citep[e.g.,][]{peim83,benj99,peim17}. The agreement between these radial trends and
the morphological patterns described in the previous section demonstrates the
consistency between the integrated and spatially resolved analyses, reinforcing
the hypothesis of a nearly spherically symmetric ionization structure.

%%%%%%%%%%%%%%%%%%%%%%%%%%%%%%%%%%%%%%%%%%%%%%%%%%%%%%%%%%%%%%%%%%%%%%%5
\begin{figure*}
  \centering
  \includegraphics[width=0.89\textwidth,clip,trim=0 0 0 0]{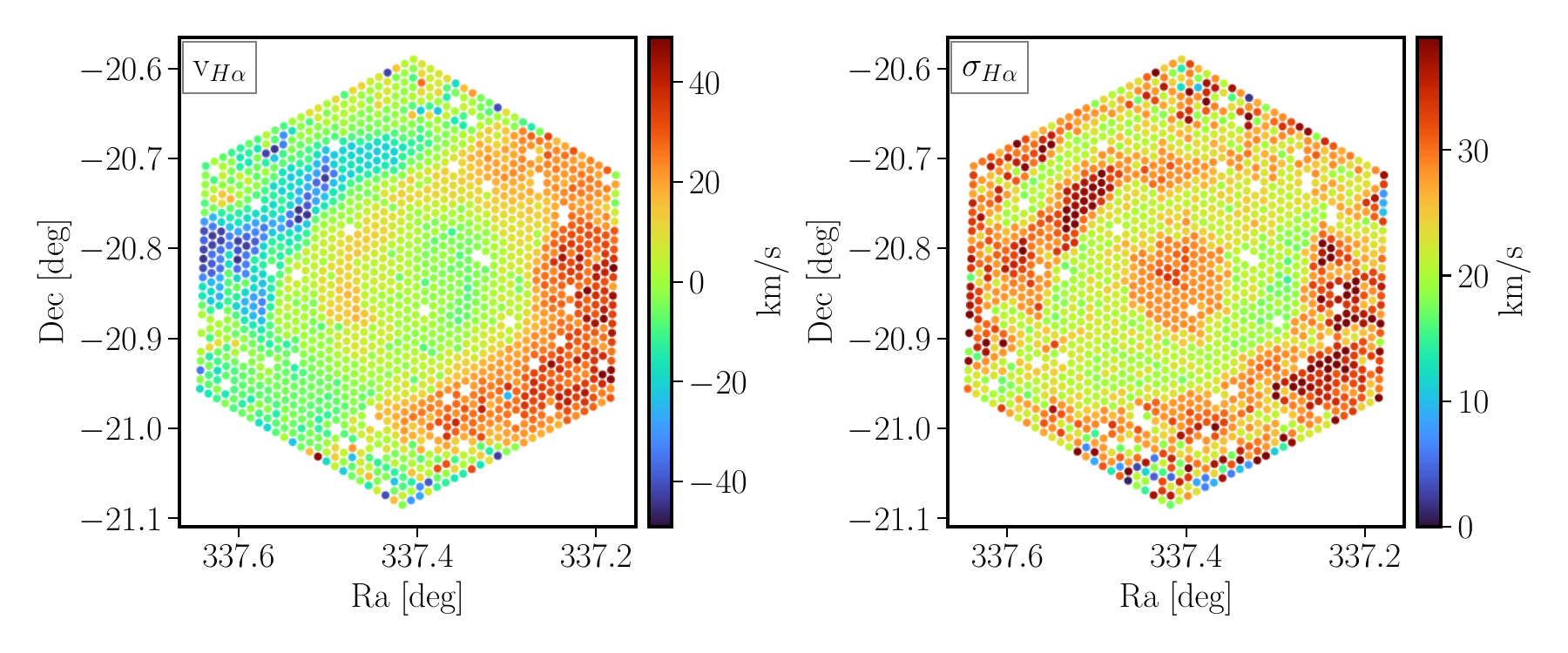}
  \captionof{figure}[]{Spatial distribution of the \ha\ velocity (left-panel) and velocity dispersion (right-panel)  derived using the 2nd parametric analysis of the emission lines by the LVM-DAP for the analyzed { Helix Nebula} dataset. The velocity map shows a blue-/red-shift pattern with a maximum absolute velocity coincident with the highest velocity dispersions, in agreement with an spherically expanding shell.}
        \label{fig:kin}
    \end{figure*}
%%%%%%%%%%%%%%%%%%%%%%%%%%%%%%%%%%%%%%%%%%%%%%%%%%%%%%%%%%%%%%%%%%%%%%%5

\subsubsection{Emission line fluxes relative to H$\beta$}
\label{sec:norm}

The right panel of Figure~\ref{fig:rad} displays the corresponding line ratios,
normalized to \Hb, as a function of radius for the same emission lines shown in
the left panel. As in the case of the fluxes themselves, the line ratios present a very small
azimuthal dispersion around the mean value at each radial distance. Again, this
supports the scenario in which the ionization present a spherical structure. 

The Balmer ratios (\ha/\hb, \hg/\hb, and \hd/\hb) remain remarkably constant as a
function of radius across the entire nebula. 
The absence of significant radial variations indicates that internal dust extinction
within the ionized shell is either extremely low or distributed very uniformly.
This behavior is consistent with earlier optical analyses of the Helix Nebula,
which also found negligible or spatially uniform reddening 
\citep[e.g.][]{odell98,henry99,meaburn05}.

Since the intrinsic Case~B Balmer ratios are only mildly sensitive to density and temperature in the typical nebular regime, the observed flatness of the radial profiles suggests that the physical conditions throughout the ionized gas do not exhibit extreme variations—of at least one order of magnitude—on the spatial scales probed by the LVM ultra-wide IFU. A detailed determination of the temperature and density structure is beyond the scope of this paper. We will address this in future work using additional pointings obtained by the LVM survey of the Helix Nebula (Orozco-Duarte et al., in prep.).

%Since the intrinsic Case~B Balmer ratios are sensitive (albeit mildly) to density and
%temperature in the typical nebular regime, the observed flatness of the radial
%profiles also suggests that the physical conditions throughout the ionized gas 
%do not present sharp variations at the scales probed by the LVM ultra-wide IFU, with
%in the sensitivity of the Balmer lines to those changes.   
%A detailed determination of the temperature and density structure is beyond the
%scope of this paper. We will address it in future explorations using additional
%pointings being acquired by the LVM survey of the { Helix Nebula} (Orozco-Duarte et al., in prep.).

Beyond the Balmer emission lines, the remaining line ratios displayed
in the right-hand panel of Fig.~\ref{fig:rad} reveal systematic
radial trends that mirror the ionization stratification already identified in the two-dimensional maps (Sec.~\ref{sec:res_map}) and in the flux profiles of the left panel. High-ionization ratios, such as [\ION{O}{iii}]5007/\Hb, exhibit a pronounced peak toward the inner regions, reflecting the concentration of O$^{++}$ in zones exposed to the hardest radiation field near the central star. This behavior is fully consistent with the compact morphology of He\,{\sc ii}\,$\lambda 4686$ and with the enhanced [\ION{O}{iii}] emission inside the main ring.  The gradual decline of these ratios with increasing radius, sharper for the He line, indicates a softening of the ionizing continuum as high-energy photons are progressively absorbed in the inner nebular layers, as described before.

Intermediate-excitation ratios, such as He\,{\sc i}\,$\lambda 6678$/\Hb\ or
[\ION{Ar}{iii}]7136/\Hb, show a broader plateau across the main ring,
reaching their maximum where the He$^{+}$ zone is most extended. Their
slow radial variation reflects the fact that these ions dominate over a
large volume of the nebula, consistent with the relatively smooth
morphology seen in their two-dimensional maps.  On the contrary, low-ionization ratios, such as [\ION{N}{ii}]6583/\Hb,
[\ION{O}{ii}]3727/\Hb, and [\ION{S}{ii}]6717,6731/\Hb, rise steeply toward
the outer nebula, peaking just beyond the [\ION{O}{iii}] bright ring.
This behavior echoes the enhancements seen in their spatial maps,
where these lines trace the edge of the ionization front and the
clump-dominated low-ionization structures. The outward increase in these
ratios reflects the transition from hydrogen-ionized gas to partially
ionized or neutral material and confirms the classical ionization
stratification of the Helix Nebula. The coincidence of their maxima
with the projected location of the ionization front agrees with earlier
studies that associate these low-ionization lines with dense knots and
filamentary structures in the outer ring, as discussed before \citep[e.g.][]{odell98,meaburn92}.

\subsection{Ionized gas kinematics}
\label{sec:res_kin}

%The resulting patterns are characteristic of an expanding
%ionized shell, consistent with the canonical picture of planetary
%nebula evolution \citep{oster06,odell98,meaburn05,odell07}.  
    
 Figure~\ref{fig:kin} shows the observed velocity\footnote{not corrected by Heliocentric velocity} ($v_{H\alpha}$) and
 velocity-dispersion ($\sigma_{H\alpha}$) maps derived for the H$\alpha$ emission line, extracted from the 2nd parametric analysis of the emission lines described in Sec. \ref{sec:ana}, tracing the bulk kinematics of the ionized gas throughout the inner 40$\arcmin$ of the Helix Nebula.  The line-of-sight velocity map reveals a smooth, large-scale gradient with a clear redshifted–blueshifted symmetry along a preferential axis.  This structure is consistent with the projection of a slowly expanding, moderately inclined shell, as previously inferred from long-slit spectroscopy and Fabry--Perot interferometry \citep{meaburn92,walshmeaburn87,odell98}.  The approaching (blueshifted) region lies to the northwest, while the receding (redshifted) region extends to the southeast, consistent with earlier determinations of the nebula's orientation and dynamical axis \citep[e.g.][]{odell98,meaburn05}.  The measured velocity amplitudes, $\sim$33 km\,s$^{-1}$, align with previous estimates for the global expansion of the ionized shell \citep{henry99,odell04}, confirming that the Helix remains a relatively low-velocity, evolved planetary nebula.

The H$\alpha$ velocity-dispersion map, once corrected by the intrinsic velocity dispersion as described in Sec. \ref{sec:ana}, shows enhanced values toward the
inner cavity and along the brightest sections of the main ring,
while lower dispersions are found in the outer regions.  These narrow
line widths (typically $\sigma_{H\alpha}\sim $10-15~km\,s$^{-1}$) are consistent
with thermal broadening at electron temperatures of
$T_{\mathrm{e}}\sim$10$^4$\,K plus a modest contribution from unresolved
turbulent motions, matching previous detailed studies
\citep{henry99,odell07}.  Slightly broader profiles in the inner
regions coincide with the line-of-sight integration of the front and
back sides of the expanding shell, an effect commonly observed in
planetary nebulae with hollow or cavity-type geometries
\citep[e.g.][]{grue92,tyle03}.

The combined velocity and dispersion maps support the interpretation
that the Helix Nebula traces the limb-brightened edges of a thick,
slowly expanding, quasi-spherical bubble\citep[e.g.][]{odell07}.  This dynamical structure is
a natural outcome of the canonical interacting stellar wind model for
planetary nebulae, in which fast winds from the hot central star sweep
up the previously ejected AGB envelope into an expanding ionized shell
\citep[e.g.][]{oster06}.  The velocity field extracted from H$\alpha$ mirrors
the signatures expected from such a geometry: redshift–blueshift
symmetry, enhanced line splitting toward the nebular center, and
moderately increasing dispersions where the shell is intersected along
longer sight-lines.  These properties are in strong agreement with
classic kinematic analyses of the Helix and other evolved nebulae
\citep[e.g.][]{odell98,meaburn05,odell07}.

The kinematical patterns recovered from the LVM data closely resemble
those reported in earlier high-resolution optical studies, including
the global expansion velocity of $\sim$20-30~km\,s$^{-1}$, the
inclination of the ring-like structure, and the mild internal
turbulence inferred from linewidths
\citep{walshmeaburn87,meaburn92,meaburn05}.  The excellent agreement
demonstrates that the wide-field LVM observations reliably capture the
large-scale dynamical structure of this prototypical planetary nebula.

%In summary, the H$\alpha$ kinematic maps confirm that the Helix Nebula
%is an evolved, photoionized shell undergoing slow, nearly symmetric
%expansion.  The observed velocity structure, dispersion patterns, and
%orientation are fully consistent with the general dynamical framework
%for planetary nebulae and match, in detail, previous determinations
%from targeted spectroscopic studies.

\subsection{Stellar population content}
\label{sec:st}    
    
%%%%%%%%%%%%%%%%%%%%%%%%%%%%%%%%%%%%%%%%%%%%%%%%%%%%%%%%%%%%%%%%%%%%%%%5
\begin{figure}
  \centering \includegraphics[width=0.47\textwidth,clip,trim=0 0 0 20]{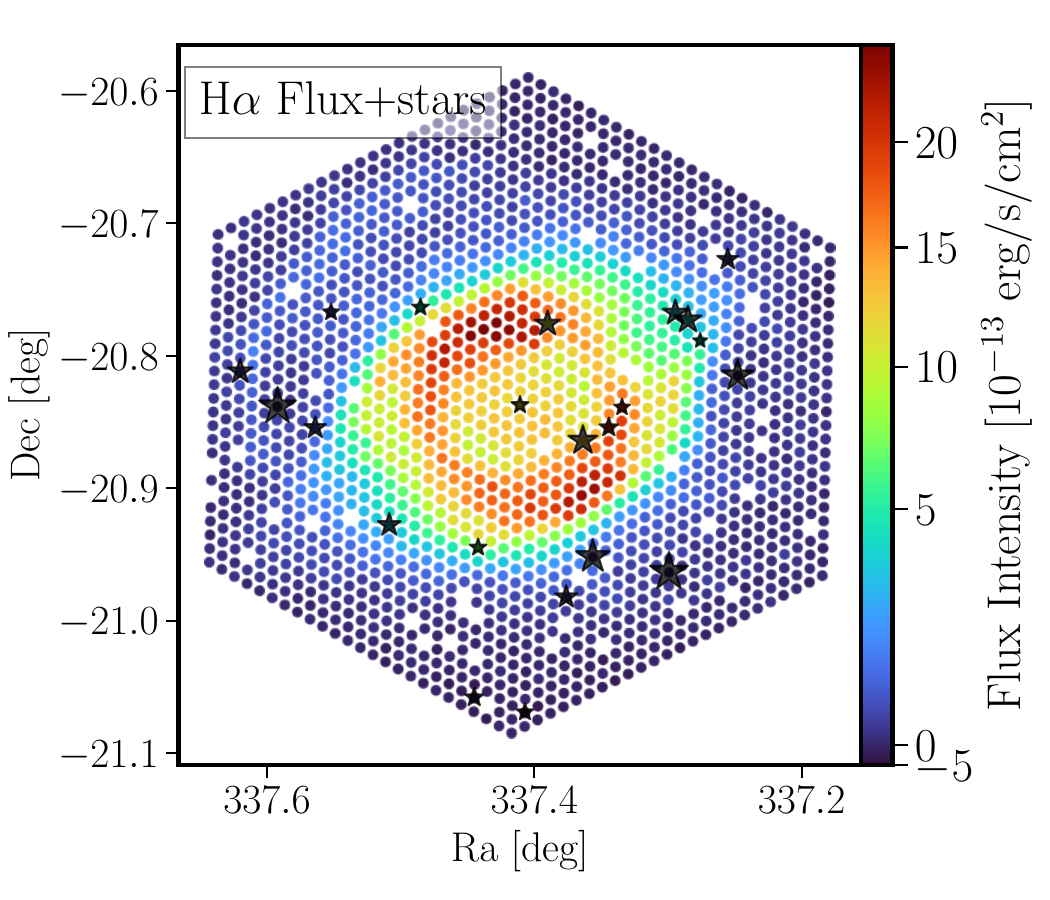}
  \captionof{figure}[]{Distribution
    across the FoV of the fibers in which the stellar spectra have a
    S/N$>$5 at $\sim$5500 (solid stars) together
    with the H$\alpha$ intensity map (already shown in
    Fig. \ref{fig:fe}) in arbitrary units, used as spatial reference.
    The size of the stars corresponds to the S/N, ranging from $\sim$5
    for the faintest ones to $\sim$30 for the brightest ones. The
    central star of the { Helix Nebula} has a S/N just above 5.}
        \label{fig:st}
    \end{figure}
    %%%%%%%%%%%%%%%%%%%%%%%%%%%%%%%%%%%%%%%%%%%%%%%%%%%%%%%%%%%%%%%%%%%%%%% 5

As summarized in Sec. \ref{sec:ana}, and described in detail in
 \citet{lvmdap}, the LVM-DAP models the stellar spectra by decomposing them into a set of RSP templates, stellar spectra characteristic of a particular distribution of physical properties that minimize the degeneracies among them. This kind of analysis provides realiable results when the signal-to-noise reaches a certain threshold that depends strongly on the number of templates considered in the decomposition. For the currently assumed RSP library, which comprises 108 templates, this minimum S/N is $\sim$40-50, based on simulations \citep{lvmdap}. Below this threshold, the procedure provides a model for the underlying stellar component, but in that case the estimated physical parameters are not completely reliable. Nevertheless, the configuration adopted for the current analysis perform sthis decomposition when the S/N$>$20. Below this limit, if S/N$>$1, the procedure just looks for the single RSP that best fits the underlying continuum. In other cases, the underlying continuum is just ignored in the analysis.

%%%%%%%%%%%%%%%%%%%%%%%%%%%%%%%%%%%%%%%%%%%%%%%%%%%%%%%%%%%%%%%%%%%%%%%5
\begin{figure}[t]
  \centering
  \includegraphics[width=0.47\textwidth,clip,trim=20 15 100 70]{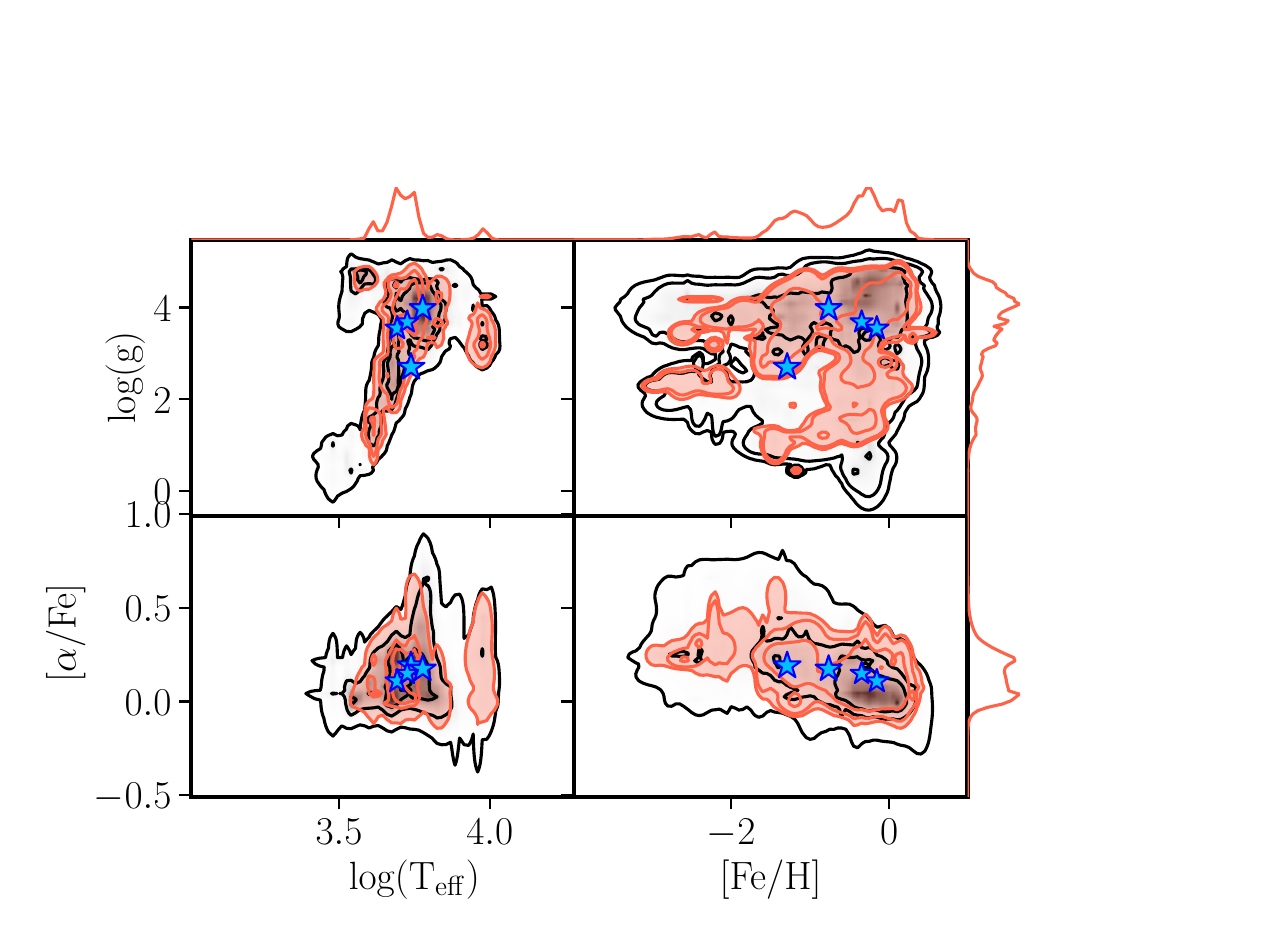}  
  \captionof{figure}[]{Probability distribution function of the physical properties of stars (T$_{eff}$, log($g$), [Fe/H] and [$\alpha$/Fe]) for the full RSP template comprising 108 RSPs (black contours), together with the same distribution for the four stars in the FoV with a S/N$>$25 (see Fig. \ref{fig:st} based on the analysis with the DAP. Each panel shows the PDFs for a pair of physical properties: T$_{eff}$-log($g$) (top-left); [Fe/H]-log($g$) (top-right); T$_{eff}$-[$\alpha$/Fe] (bottom-left) and [$\alpha$/Fe]-[Fe/H] (bottom-right).  In each panel each successive contour corresponds approximately to 1, 2, and 3$\sigma$. The histograms at the top- and right-hand panels show the projected PDFs for each individual parameter. The luminosity-weighted values for each of the four stars are represented by blue stars.}
        \label{fig:rsp}
    \end{figure}
    %%%%%%%%%%%%%%%%%%%%%%%%%%%%%%%%%%%%%%%%%%%%%%%%%%%%%%%%%%%%%%%%%%%%%%% 5

The  LVM exposure analyzed along this study, centered on the { Helix Nebula}, was acquired during
the science commissioning explorations with the main aim of exploring the ionized ISM, and it was never
intended to obtain high-quality S/N spectra of the stellar content in the (projected) vicinity
of the nebula. Thus, neither the central ionizing star nor the field stars present a particularly
high S/N. Figure \ref{fig:st} shows the distribution within the FoV of the 22 fibers in which
the stellar spectra have a S/N$>$5 at 5000\AA\ for the analyzed dataset. Only 4 of them present
a S/N$>$25, and none of them are sufficiently bright to provide a reliable decomposition of
the stellar content. In particular, the fiber covering the central star do present a low S/N, $\sim$6.

Nevertheless, as the main purpose of the current exploration is to provide a showcase of the
analysis provided by the LVM-DAP, describing the content of the dataproducts presented in Sec. \ref{sec:res_dp}, we present here the results of the stellar decomposition for the 4 fibers with higher S/N. Figure \ref{fig:rsp} shows the PDF of the stellar properties derived combining
the results of the DAP decomposition for these spectra, projected in the log(g)-Teff, log(g)-[Fe/H], [$\alpha$/Fe]-Teff and [$\alpha$/Fe]-[Fe/H] planes, compared with the distribution for the full MaStar stellar library \citep[as described in ][]{lvmdap}. The luminosity weighted average
values for the four fibers are included for comparison purposes. We should stress that this stellar component
corresponds to field stars that we cannot confirm to be related in any way with the studied nebula, and that the results, due
to limited S/N are not reliable. Therefore, we refrain from making any discussion or extracting any conclusion
on the presented results. For an example of a direct connection between the stellar component and the ISM observed by
the LVM, we refer the reader to \citet{villa25}. In this recent study, authors illustrate the power of these data unveiling the spatially resolved interplay between ionized gas, molecular material, dust  and stellar content in the Rosette Nebula across large angular scales.

\section{DISCUSSION}
\label{sec:dis}

%%%%%%%%%%%%%%%%%%%%%%%%%%%%%%%%%%%%%%%%%%%%%%%%%%%%%%%%%%%%%%%%%%%%%%%5
\begin{figure*}[t]
        \centering                
        \includegraphics[width=8.5cm,clip,trim=0 0 0 0]{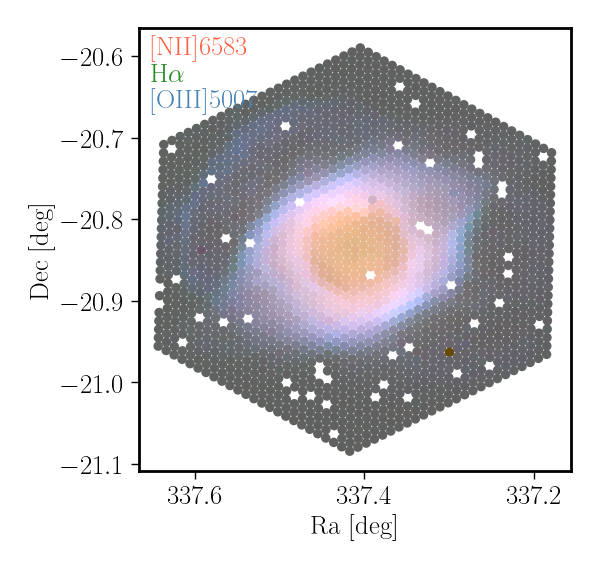}
        \includegraphics[width=8.5cm,clip,trim=0 0 0 0]{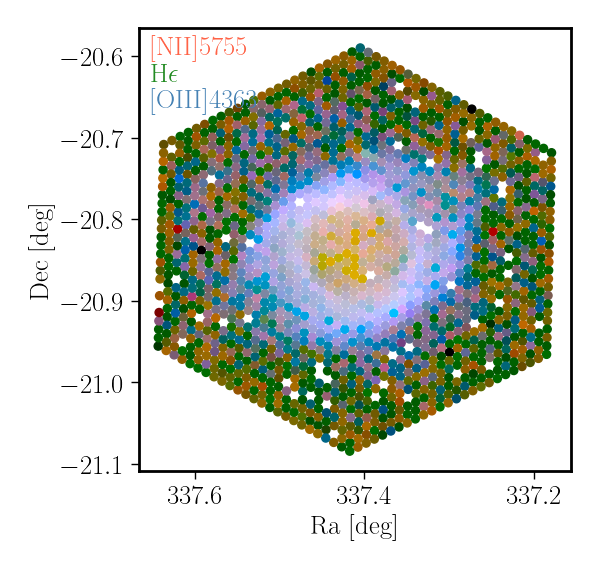}
        \captionof{figure}[]{{\it Left-panel:} Distribution of the fluxes recovered by the DAP for the \oiii~$\lambda5007$ (blue), H$\alpha$ (green), and \nii~$\lambda6583$ (red) strong emission lines using the non-parametric analysis for the analyzed pointing on the { Helix Nebula}. {\it Right-panel:} Similar distribution as the one shown in the right-panel for the \oiii~$\lambda4363$ (blue), H$\epsilon$ (green), and \nii~$\lambda5755$ (red) weak emission lines. The fluxes are displayed using arbitrary intensity scale chosen to enhance contrast, using the same scaling for the three emission lines considered in each figure.}
        \label{fig:rgb}
      \end{figure*}
      %%%%%%%%%%%%%%%%%%%%%%%%%%%%%%%%%%%%%%%%%%%%%%%%%%%%%%%%%%%%%%%%%%%%%%% 5

%----------------------------------------EL HELIX NP-----------------------
\begin{table*}[ht]
\RaggedRight
\caption{Emission line fluxes across the { Helix Nebula}e: comparison with the literature}
\label{tab:comp}
\begin{tabular}{lcccccccc}
\toprule  
 Emission Line ($\lambda$) & \multicolumn{4}{c}{\citet{odell98}} & \multicolumn{3}{c}{\citet{henry99}} & LVM pointing\\
   (\,\AA) & (1) & (2) & (3) & (4) & (A) & (B) & (C) & (Integrated) \\
\midrule
{[O\,\textsc{ii}]} 3727 & 103 & 143 & 328 & 674 & 188 & 518 & 718 &  454.8 \\
He\,\textsc{ii}\,4686  & 61.2 & 45.6 & 20.3 & 4.6 & 8.9 & 3.3 & 0.46 &   10.4 \\
{[O\,\textsc{iii}]} 4959 & 259 & 293 & 284 & 190 & 209 & 167 & 81 &   155.0 \\
{[O\,\textsc{iii}]} 5007 & 754 & 841 & 818 & 551 & 714 & 507 & 309 &   496.2 \\
{[N\,\textsc{ii}]} 6548 & 40.4 & 33.1 & 77.1 & 186 & 81 & 150 & 280 &  156.1 \\
H$\alpha$\,6563 & 270 & 281 & 278 & 288 & 272 & 286 & 284 &  277.1 \\
{[N\,\textsc{ii}]} 6583 & 74.1 & 100 & 243 & 575 & 250 & 465 & 842 &  470.4 \\
{[S\,\textsc{ii}]} 6717 & 3.1 & 4.0 & 4.9 & 12.6 & 3.6 & 9.2 & 22 &  16.9 \\
{[S\,\textsc{ii}]} 6731 & 2.1 & 2.7 & 3.4 & 8.9 & 2.6 & 6.6 & 16 &  12.2 \\
{[O\,\textsc{i}]} 6300 & 2.6 & 6.3 & 9.1 & 22.5 &  &  &  &  35.9$^1$ \\
{[O\,\textsc{i}]} 6363 & 0.8 & 2.8 & 3.3 & 7.9 &  &  &  &  11.9$^1$ \\
He\,\textsc{i}\,5876 & 10.0 & 12.5 & 14.7 & 18.2 & 15 & 17 & 17 &  19.3 \\
\bottomrule
\end{tabular}    
\tabletext{Fluxes are normalized to H$\beta$=100 10$^{-13}$ erg/s/cm$^2$. (1) Emission lines possible
affected by imperfections in the sky subtraction, as discussed in the text.}
\end{table*}
%----------------------------------------------------------------

The single exposure taken by the LVM of the Helix Nebula distributed
in the SDSS DR19 provides one of the deepest and most spatially
complete optical spectroscopic data sets ever obtained for this
object. A particularly compelling demonstration of the capabilities of
the survey is the detection of faint auroral and
high-excitation diagnostic lines across a very large field of
view. Figure~\ref{fig:rgb} illustrates this clearly, showing a direct
 comparison between the spatial distribution of bright nebular lines such as [\ION{O}{iii}]5007, \ha , [\ION{N}{ii}]6583, and the much weaker counterparts [\ION{N}{ii}]5755, \he\ and [\ION{O}{iii}]4363.  These lines typically reach intensities of $\sim$0.5-2\% of \Hb\ in classical long-slit observations, and in some cases even lower, making them notoriously difficult to measure reliably over extended regions. Nevertheless, it is evident that both bright and weak lines trace a similar distribution within the nebula.

Fig.~\ref{fig:rgb} shows that both [\ION{O}{iii}]4363 and [\ION{N}{ii}]5755 are coherently detected even in regions where the surface brightness drops by several orders of magnitude relative to the bright central ring for the strongest lines (e.g. [\ION{O}{iii}]5007). This spatial robustness, not achievable with earlier slit- or aperture-based studies, is essential for constraining temperature variations and potential small-scale thermal inhomogeneities throughout the nebula \citep[e.g., ][Méndez-Delgado et al., in prep.; Egorov et al. in prep.; Singh et al. in prep.]{kreck25,satt25,sarb25}.The recovery of such weak features across more than thousands of independent fibers results from the combination of wide-field IFU coverage ($\sim$0.45\textdegree$^2$), the size of the fibers projected in the sky ($\sim$35.3$\arcsec$), stable spectrophotometric calibration, and accurate sky subtraction intrinsic to the LVM instrument and survey design and the capabilities of the DRP. This is achieved despite still remaining imperfections of the sky subtraction, as discussed in Sec. \ref{sec:ana}. The ongoing improvements in 
the development of the DRP suggest not only the resolution of this issue, but a clear improvement in the final depth achieved by the very same observations (Mej\'\i a-Narv\'aez et al. in prep.).

As indicated along the manuscript several previous observations have already explored the { Helix Nebula}, using narrow-band imaging and focused slit-spectroscopy. Table~\ref{tab:comp} provides a quantitative comparison between the LVM line ratios and a subset of those published in the classical studies \citep[e.g.][]{odell98,henry99}. We should recall that the exact location from which those values were extracted was already
shown in Fig. \ref{fig:wise}. Overall, the LVM values fall well within the range spanned by previous measurements, despite the large diversity of apertures, slit widths, and extraction geometries employed in the historical literature, in particular when the spatial (radial) variations of the different line fluxes (and ratios), are taken into account (e.g., Fig. \ref{fig:fe} and \ref{fig:rad}). For example, the LVM [\ION{O}{iii}]5007/\Hb\ ratio lies between the inner and outer apertures reported by \citet{odell98}, consistent with the fact that the LVM footprint covers both high- and intermediate-ionization zones projected on the sky. Similarly, the intensities measured for [\ION{O}{iii}]4363 and [\ION{N}{ii}]5755 agree with earlier determinations by \citet{leene87} and \citet{henry99}, validating the accuracy of the LVM flux calibration even at levels 100-1000 times fainter than the \hb\ level. 

Where differences appear, they are clearly attributable to spatial variations within the nebula and to differences in the apertures and their locations among the classical studies. Earlier spectroscopic measurements were obtained through narrow slits or small circular apertures, sampling either the inner cavity or the bright ring, whereas the LVM ratios represent area-averaged values over a large, contiguous hexagonal footprint. Because the Helix exhibits strong ionization stratification, integrated values will naturally differ from those derived from discrete, spatially limited regions. The systematic behavior observed in Fig. \ref{fig:rad}, and the comparisons included in Fig. \ref{fig:rgb} and Tab. \ref{tab:comp}, reinforce this interpretation: lines arising from high-excitation species (e.g. He,\textsc{ii},4686, [\ION{O}{iii}],4363) present lower integrated/average values across
the nebula (i.e., LVM values) than in the inner-aperture measurements of \citet{odell98}, whereas low-ionization features (e.g. [\ION{N}{ii}],5755, [\ION{O}{i}],6300) are enhanced compared to values obtained from apertures centered on the cavity. This trend is precisely what is expected for a geometrically thick, nearly face-on planetary nebula dominated by an ionization-bounded main ring, which is the general scenario described along this study.

The combined results therefore demonstrate two fundamental outcomes of the LVM observations: (i) the data reach a depth and spatial uniformity that allow weak diagnostic lines to be mapped across the full nebular extent with high fidelity, something unattainable in previous optical spectroscopic studies of NGC~7293; and (ii), the consistency of the LVM line ratios with historical measurements, once aperture effects are accounted for, validates the accuracy of the LVM-DRP and confirms that the Helix Nebula's canonical ionization structure as inferred from decades of long-slit and narrow-band work is fully reproduced by the new dataset. These results highlight the LVM's unique capability to combine spectroscopic depth with true wide-field coverage, enabling a comprehensive spatial characterization of nebular excitation, temperature, and density conditions in a manner not previously possible.

A further strength of this analysis lies in the performance of the LVM Data Analysis Pipeline (LVM-DAP), which is specifically optimized to recover both strong and extremely weak emission features across tens of thousands of independent spectra. The combination of robust stellar-continuum subtraction, iterative Monte Carlo error propagation, and dual parametric and non-parametric emission-line measurements ensures that faint diagnostic lines, such as [\ION{O}{iii}]~4363, [\ION{N}{ii}]~5755, and high-order Balmer lines, are extracted with statistically meaningful uncertainties even at very low equivalent widths, as predicted in \citet{lvmdap}.These capabilities, combined with the excellent homogeneity of the spectrophotometric calibration  provided by the LVM instrument, demonstrate that the DAP is able to reproduce the classical nebular diagnostics, and improves, significantly, the spatial completeness and sensitivity compared to previous studies.

\section{CONCLUSIONS}
\label{sec:con}

In this work we have presented a detailed spectroscopic analysis of a single SDSS-V Local Volume Mapper (LVM) pointing of the Helix Nebula (NGC~7293), obtained during science commissioning. Despite being acquired in an early phase of operations, and reduced with a pipeline still under development (Mejía-Narváez et al., in prep.), the data demonstrate the remarkable sensitivity, spatial completeness, and spectrophotometric stability of the LVM system, enabling a comprehensive view of the ionized-gas structure across the bright ring, central cavity, and inner halo of one of the nearest planetary nebulae.

The spatial distributions of the main nebular emission lines reveal the expected ionization stratification of a photoionized, moderately inclined shell: high-excitation species such as [\ION{O}{iii}]5007 peak in the inner portions of the main ring, intermediate-ionization tracers such as He,\textsc{i},6678 follow closely the morphology of the Balmer emission, and low-ionization lines such as [\ION{N}{ii}]6583, [\ION{S}{ii}]6717,6731, and [\ION{O}{ii}]3727 delineate the outer ionization front and associated knots and filaments. These results are fully consistent with classical narrow-band imaging \citep[e.g.][]{odell98,meaburn05,odell07} and provide the first fiber-by-fiber spectroscopic mapping across the full nebular face.

The radial analysis reinforces the results derived from the 2D maps, with azimuthally averaged flux profiles showing that each ionic species peaks at the radius expected from its ionization potential. Balmer lines remain remarkably constant in their ratios relative to H$\beta$, indicating that dust extinction is extremely low and spatially uniform, and that the nebular electron temperature and density vary only modestly across the observed field. These results echo previous slit-based determinations of the homogeneous physical conditions in the Helix \citep[e.g.][]{henry99} while offering a panoramic view not accessible to earlier observations.

One of the major capabilities highlighted by this study is the ability of the LVM to detect very faint auroral lines across the entire field.  The detection of  temperature-sensitive auroral lines on a fiber-by-fiber basis is unprecedented for a Galactic planetary nebula at this angular scale. The excellent agreement between our integrated line ratios and those from earlier long-slit studies \citep{odell98,henry99} validates both the data quality and the performance of the LVM-DAP.

The ionized gas kinematics further support the classical interpretation of the Helix as a slowly expanding, quasi-spherical shell whose limb-brightened geometry reproduces both the velocity pattern and line-width distribution observed across the field. The global kinematic signature agrees with previous Fabry-Perot and high-resolution spectroscopic studies \citep[e.g.][]{meaburn05}, demonstrating that even in its commissioning stage, the LVM delivers reliable velocity information at the native spatial sampling.

Taken together, these results establish the LVM as a transformative instrument for the spectroscopic study of nearby extended nebulae. The combination of large areal coverage, uniform spectrophotometry, and sensitive DAP extraction enables simultaneous access to bright diagnostics, faint auroral lines, and spatially resolved kinematics across tens of thousands of sightlines. The present analysis of the Helix Nebula serves as a proof of concept for the scientific return expected from the full LVM survey, which will provide an unprecedented census of ionized-gas structures throughout the Milky Way and Local Volume.

\section{ACKNOWLEDGEMENTS}

SFS acknowledges the support by CBF-2025-I-236 project granted by the Secretaría de Ciencia, Humanidades, Tecnología e Innovación (SECIHTI) of
the Mexican Federal Government, and the PID2022-136598NB-C31 (ESTALLIDOS) grant by the Spanish Ministery of Science and Innovation (MCINN).
JEMD, CM, SFS, ROD, LS, JT and CRZ thank the support by SECIHTI CBF-2025-I-2048 project ``Resolviendo la Física Interna de las Galaxias: De las Escalas Locales a la Estructura Global con el SDSS-V Local Volume Mapper'' (PI: M\'endez Delgado). G.A.B. acknowledges the support from the ANID Basal project FB210003. 
J.G.F-T gratefully acknowledges the grants support provided by ANID Fondecyt Postdoc No. 3230001 (Sponsoring researcher), the Joint Committee ESO-Government of Chile under the agreement 2023 ORP 062/2023, and the support of the Doctoral Program in Artificial Intelligence, DISC-UCN.
KK gratefully acknowledges funding from the Deutsche Forschungsgemeinschaft (DFG, German Research Foundation) in the form of an Emmy Noether Research Group (grant number KR4598/2-1, PI Kreckel) and the European Research Council’s starting grant ERC StG-101077573 (“ISM-METALS"). OE acknowledges funding from the Deutsche Forschungsgemeinschaft (DFG, German Research Foundation) -- project-ID 541068876.

Funding for the Sloan Digital Sky Survey V has been provided by the Alfred P. Sloan Foundation, the Heising-Simons Foundation, the National Science Foundation, and the Participating Institutions. SDSS acknowledges support and resources from the Center for High-Performance Computing at the University of Utah. SDSS telescopes are located at Apache Point Observatory, funded by the Astrophysical Research Consortium and operated by New Mexico State University, and at Las Campanas Observatory, operated by the Carnegie Institution for Science. The SDSS web site is \url{www.sdss.org}.

SDSS is managed by the Astrophysical Research Consortium for the Participating Institutions of the SDSS Collaboration, including the Carnegie Institution for Science, Chilean National Time Allocation Committee (CNTAC) ratified researchers, Caltech, the Gotham Participation Group, Harvard University, Heidelberg University, The Flatiron Institute, The Johns Hopkins University, L'Ecole polytechnique f\'{e}d\'{e}rale de Lausanne (EPFL), Leibniz-Institut f\"{u}r Astrophysik Potsdam (AIP), Max-Planck-Institut f\"{u}r Astronomie (MPIA Heidelberg), Max-Planck-Institut f\"{u}r Extraterrestrische Physik (MPE), Nanjing University, National Astronomical Observatories of China (NAOC), New Mexico State University, The Ohio State University, Pennsylvania State University, Smithsonian Astrophysical Observatory, Space Telescope Science Institute (STScI), the Stellar Astrophysics Participation Group, Universidad Nacional Aut\'{o}noma de M\'{e}xico, University of Arizona, University of Colorado Boulder, University of Illinois at Urbana-Champaign, University of Toronto, University of Utah, University of Virginia, Yale University, and Yunnan University.

\section{APPENDICES}
\label{sec:app}

%TBW

\subsection{Effects of the 2nd order sky correction}
\label{app:cor}

%%%%%%%%%%%%%%%%%%%%%%%%%%%%%%%%%%%%%%%%%%%%%%%%%%%%%%%%%%%%%%%%%%%%%%%5
\begin{figure}[t]
  \centering
  \includegraphics[width=0.47\textwidth,clip,trim=0 0 0 0]{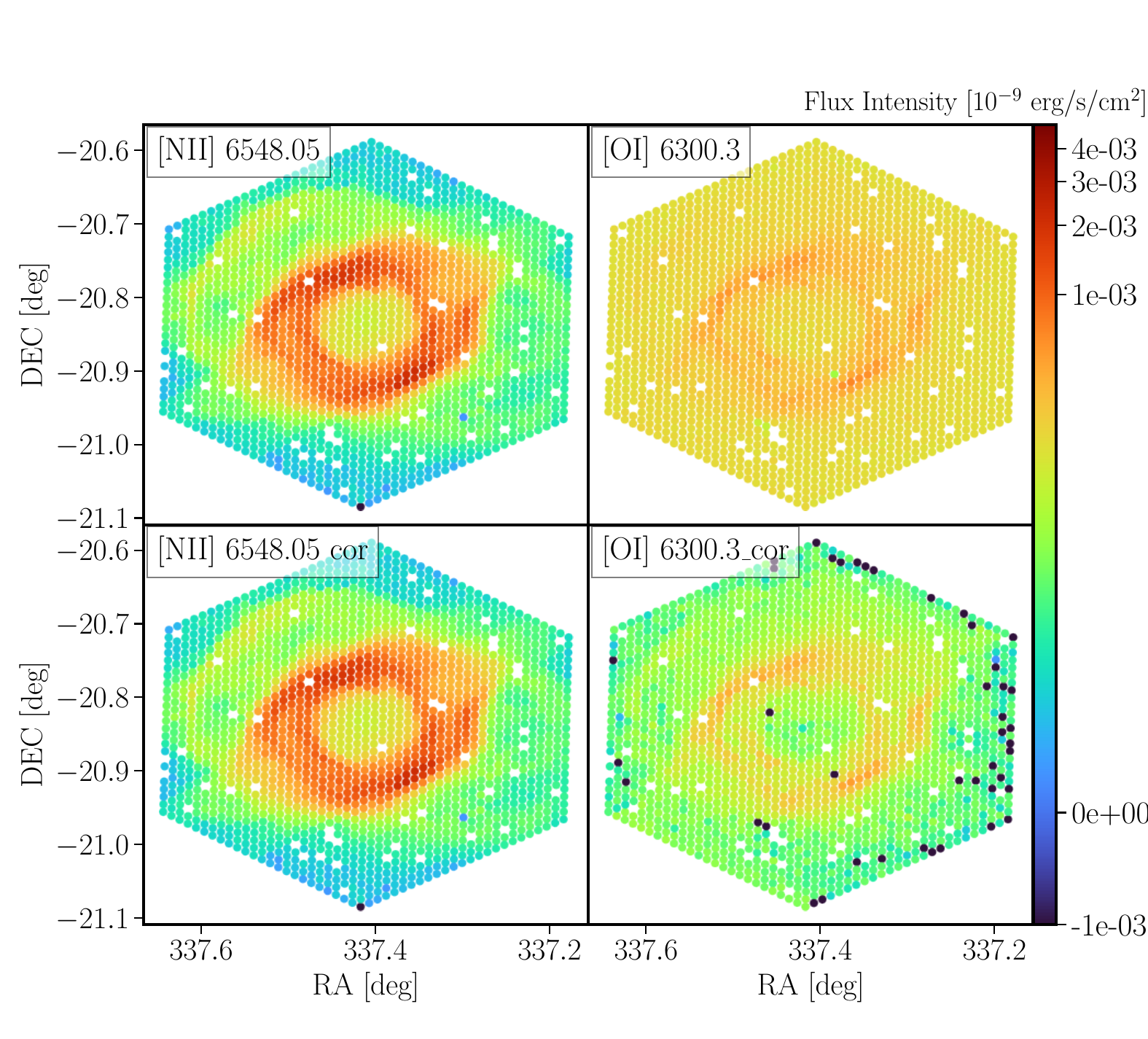}  
  \captionof{figure}[]{Comparison between the spatial distribution of the flux intensities for the [\ION{N}{ii}]6548 (left panels) and [\ION{O}{I}]6300 (right panels) emission lines before (top panels) and after (bottom panels) applying the 2nd order sky correction. While the correction has no effect in e [\ION{N}{ii}]6548, it produces a clear improvement for [\ION{O}{I}]6300.}
    \label{fig:sky}
    \end{figure}
    %%%%%%%%%%%%%%%%%%%%%%%%%%%%%%%%%%%%%%%%%%%%%%%%%%%%%%%%%%%%%%%%%%%%%%% 5

As described in Sec. \ref{sec:sky} the imperfect sky-subtraction performed by version 1.1.1
of the DRP produces an overestimation of the flux intensities for a handful of emission lines
for which we are required to perform an {\it ad hoc} 2nd order sky correction. Figure \ref{fig:sky} illustrates this effect on the spatial distribution of the flux intensities for two different emission lines: 
[\ION{N}{ii}]6548, which is barely or not affected,  and [\ION{O}{I}]6300, which is clearly affected. 
The maps on the top panels are extracted from the original DAP file \footnote{\url{https://ifs.astroscu.unam.mx/LVM_DR19_Helix/Helix_DR19_new.dap.fits.gz}}, while those
in the bottom panels are extracted from the file storing the corrected flux intensities \footnote{\url{https://ifs.astroscu.unam.mx/LVM_DR19_Helix/Helix_DR19_cor.fits.gz}}. Like the rest of the
figures included in this manuscript, this figure was created using the notebooks included
in the {\sc github} repository\footnote{\url{https://github.com/sfsanchez72/Helix_DR19}}.

The imperfect sky subtraction in the upper-left panel produces a high background intensity that creates
an apparently uniform plateau in the outer regions of the FoV. A visual inspection of the uncorrected spatial distributions for the [\ION{N}{ii}]6548 and [\ION{O}{I}]6300 flux intensities suggests a background excess in the 2nd emission line. This is confirmed by the exploration of the relative line ratios with respect to \hb, which provide nonphysical values for [\ION{O}{I}]6300, much larger than the values previously reported in the literature (e.g., Tab. \ref{tab:comp}). On the contrary, the distribution of the corrected flux intensities follows the expected pattern, with values of the same order of those reported in the literature. We should stress that (i) this contamination by a non perfect sky subtraction is affecting a very limited number of emission lines that lie very nearby (or exactly at the same wavelength) of well-known strong sky emission lines and (ii) the source of this problem has been identified and corrected in the currently developing version 1.2.0 of the DRP. Thus, we do not expect this effect to be present in future data released by the LVM survey.

\subsection{Emission line intensity maps}
\label{app:fe}

%%%%%%%%%%%%%%%%%%%%%%%%%%%%%%%%%%%%%%%%%%%%%%%%%%%%%%%%%%%%%%%%%%%%%%%5
\begin{figure*}[t]
  \centering
  \includegraphics[width=0.89\textwidth,clip,trim=30 120 0 200]{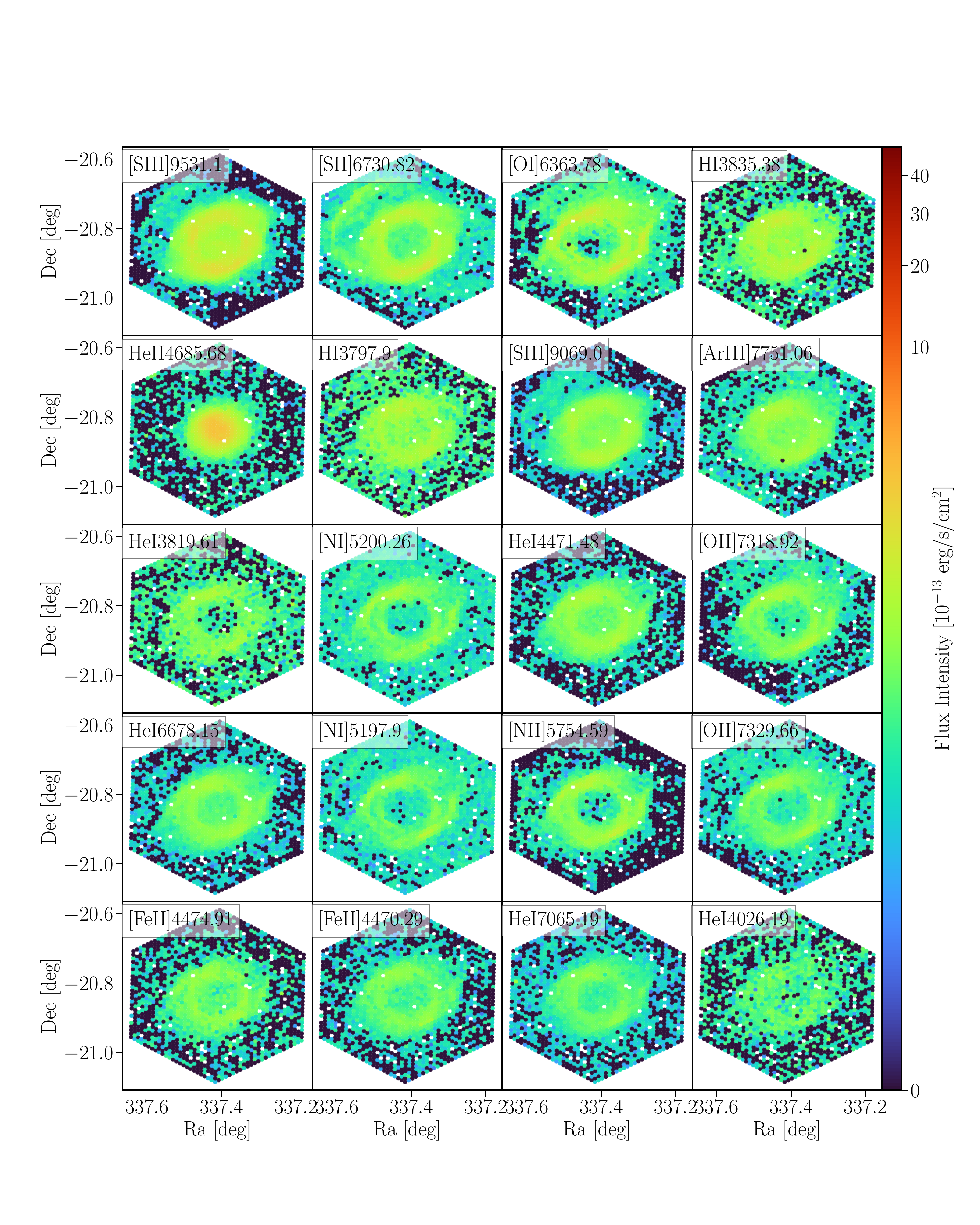}  
   \captionof{figure}[]{Similar figure as Fig. \ref{fig:fe}, for the subsequent 20 emission lines, ordered by integrated flux intensity.}
        \label{fig:fe1}
    \end{figure*}
%%%%%%%%%%%%%%%%%%%%%%%%%%%%%%%%%%%%%%%%%%%%%%%%%%%%%%%%%%%%%%%%%%%%%%% 5

%%%%%%%%%%%%%%%%%%%%%%%%%%%%%%%%%%%%%%%%%%%%%%%%%%%%%%%%%%%%%%%%%%%%%%%5
\begin{figure*}[t]
  \centering
  \includegraphics[width=0.89\textwidth,clip,trim=30 120 0 200]{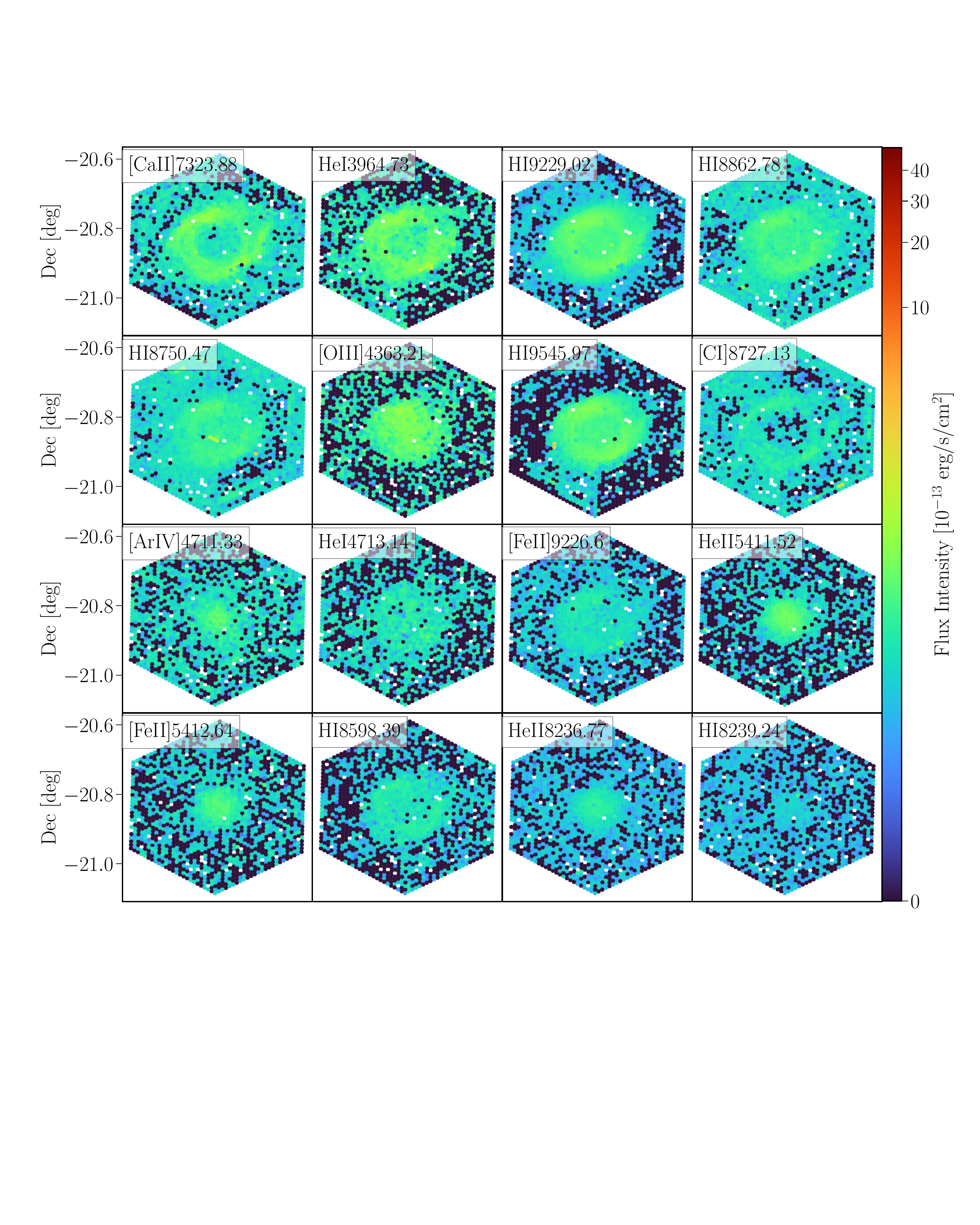}  
  \captionof{figure}[]{Similar figure as Fig. \ref{fig:fe} and \ref{fig:fe1}, for the subsequent 20 emission lines, ordered by integrated flux intensity.}
    \label{fig:fe2}
    \end{figure*}
%%%%%%%%%%%%%%%%%%%%%%%%%%%%%%%%%%%%%%%%%%%%%%%%%%%%%%%%%%%%%%%%%%%%%%%5

As already indicated in Sec. \ref{sec:res_map}, the flux intensities
for the 20 brightest emission lines included in the {\it golden sample}
are shown in Fig. \ref{fig:fe}. Figures \ref{fig:fe1} and \ref{fig:fe2}
show the spatial distribution of the flux intensities for the remaining 36
emission lines in descending order of the integrated flux intensity.
As discussed along the text, the observed spatial distributions reveal the
ionization stratification of the { Helix Nebula}: (i) strong-ionization lines
trace mostly the annular ring, being more peaked at the inner regions, (ii) low-ionization lines 
are distributed along larger radial distances, following a shallower gradient, and
(iii) emission lines that require larger excitation energies (e.g., He{\sc II} or [\ION{Fe}{II}])
are confined inside the nebular ring, with a sharp intensity drop.

\section{SOFTWARE DISTRIBUTION}
\label{app:soft}

In order to facilitate the use the dataproducts delivered by the DAP,
and in particular those delivered in the current exploration, we
deliver the python notebooks used to create all the figures shown
along this manuscript \footnote{\url{https://github.com/sfsanchez72/Helix_DR19}}.
Details on how to read the entire DAP file into a single {\tt astropy} Table,
how to plot the spatial distribution of any of the delivered parameters,
how to compare between different procedures adopted to deliver the same
parameter or/and how to handle the RSP PDF are all included as comments
within the code.

\section{DATA DISTRIBUTION}
\label{app:dist}

Along this study we made use of the reduced LVM observation on the Helix
nebula distributed as part of the SDSS-V DR19 \footnote{\url{https://dr19.sdss.org/sas/dr19/spectro/lvm/redux/1.1.1/0011XX/11111/60191/lvmSFrame-00004297.fits}}. As a result of the analysis performed by the LVM-DAP we distribute
the following dataproducts:

\begin{itemize}
\item DAP file comprising the main results of the analysis as described in Sec. \ref{sec:ana} and \citet{lvmdap}: \url{https://ifs.astroscu.unam.mx/LVM_DR19_Helix/Helix_DR19_new.dap.fits.gz}
\item Spatial distribution of the flux intensities once applied the 2nd order sky subtraction described in Sec. \ref{sec:sky}: \url{https://ifs.astroscu.unam.mx/LVM_DR19_Helix/Helix_DR19_cor.fits.gz}
\end{itemize}

\section{FACILITIES}
\label{sec:fac}

This study made use of the LVM facility, an infrastructure composed of four telescopes connected to three
spectrographs installed as Las Campanas observatory, operated by the SDSS-V collaboration.

% For observational research, authors must include a brief list of facilities and instruments used, as well as proper acknowledgment of public catalogs and virtual observatory resources.

\renewcommand{\refname}{REFERENCES}
\bibliography{my_bib}

%----------------------------------------------------------

  \end{document}